\documentclass[preprint,12pt,showpacs,aps,amsmath,amssymb, 
nofootinbib,superscriptaddress,showkeys]{revtex4} 

\usepackage[toc,page]{appendix}

\usepackage{epsfig} 
\usepackage{wasysym} 
\usepackage{mathrsfs} 
\usepackage{graphicx} 
\usepackage{amsfonts} 
\usepackage{amsbsy} 
\usepackage{amscd} 
\usepackage{pstricks} 

\newcommand{\be}{\begin{equation}}
\newcommand{\ee}{\end{equation}}
\newcommand{\bes}{\begin{equation*}}
\newcommand{\ees}{\end{equation*}}
\newcommand{\bea}{\begin{eqnarray}}
\newcommand{\eea}{\end{eqnarray}}
\newcommand{\beas}{\begin{eqnarray*}}
\newcommand{\eeas}{\end{eqnarray*}}
\newcommand{\Tr}{\text{Tr}}
\newcommand{\la}{\lambda}
\begin{document}

\author{Joydeep Chakrabortty}
\affiliation{Department of Physics, Indian Institute of Technology, Kanpur-208016, India}

\author{Partha Konar}
\email{konar@prl.res.in}
\affiliation{Physical Research Laboratory, Ahmedabad-380009, India}

\author{Tanmoy Mondal}
\email{tanmoym@prl.res.in}
\affiliation{Physical Research Laboratory, Ahmedabad-380009, India}

\title{Constraining a class of $B-L$ extended models from vacuum stability and perturbativity}

\begin{abstract}
The precise knowledge of the Standard Model (SM) Higgs boson and top--quark masses and couplings are crucial to 
understand the physics beyond it. An SM--like Higgs boson having a mass in the range of 123--127 GeV squeezes the parameters 
for physics beyond the Standard Model. In recent the LHC era many TeV--scale neutrino mass models have earned much attention as they pose 
many interesting phenomenological aspects. We have contemplated $B-L$ extended models which are theoretically well 
motivated and phenomenologically interesting, and they successfully explain neutrino mass generation. 
In this article we analyze the detailed structures of the scalar potentials
for such models. We compute the criteria which guarantee that the vacuum is bounded from below in all directions. 
In addition perturbativity (triviality) bounds are also 
necessitated. Incorporating all such effects we constrain the parameters of such models by performing their renormalization group evolutions. 
\end{abstract}



\maketitle

\section{Introduction} \label{sec:intro}
 
The recent announcements from both ATLAS \cite{Aad:2012tfa} and CMS \cite{Chatrchyan:2012ufa} have revealed 
the existence of a new boson having a mass in the range 123--127 GeV. The data so far indicates a 
close resemblance to one having some of the measured properties of the Standard Model (SM) Higgs. However, it has yet to   
confirm firmly whether this boson is the SM Higgs or a beyond the Standard Model artifact. 
This long awaited quest will only be examined more vigorously in the near future with the help of more data. 
 
If the newly discovered particle is indeed the SM Higgs boson then its mass can carry a signature of new physics which embeds SM at  low energy. 
The Higgs mass can be recast solely in terms of the Higgs quartic coupling, $\lambda_h$. The stability of the 
electroweak (EW) vacuum demands a positive $\lambda_h$. Now if  the  SM is the only existing theory in nature then this condition, $\lambda_h>0$\footnote{This is the necessary
condition but not the sufficient to confirm the sole existence of the SM till the Planck scale.}, 
must be maintained at each scale of its evolution up to the Planck scale ($M_{Pl}$). The evolution 
of $\lambda_h$ with the  renormalization (mass) scale limits two boundary values -- one  at the EW scale for which we have 
$\lambda_h(M_{Pl})=\pi$, and one at the Planck scale for which we have 0 -- from the demands of perturbativity of the coupling (triviality)
and the stability of the vacuum (vacuum stability) respectively. It has been noted in Refs.~\cite{Degrassi:2012ry_vs,Alekhin:2012py_vs,Masina:2012tz_vs}
that the SM electroweak vacuum is not stable up-to the Planck scale for most of the SM parameters (top-quark mass, Higgs mass and strong coupling $\alpha_s$).
Thus it indicates that some new physics might be there before the SM vacuum stability gets raptured. Thus the physics beyond Standard Model is expected to 
take care of stability of the vacuum of the full scalar potential along with the electroweak ones. In brief, the present range of the SM-like Higgs mass
entertains the presence of new physics solely from the vacuum stability point of view. 

Apart from this, we already have hints of new physics beyond the 
Standard Model from the neutrino sector. Many experimental observations, like neutrino oscillations, confirm that neutrinos have tiny nonzero 
masses which cannot be accommodated naturally within the SM. Thus we must have physics beyond the Standard Model to explain  
this feature. Among the neutrino mass generation procedures the  seesaw mechanism
\cite{Minkowski:1977sc_seesawI,Yanagida:1979as_seesawI,GellMann:1980vs_seesawI,Glashow:1979nm_seesawI,Mohapatra:1979ia_seesawI,Magg:1980ut_seesawII,
Lazarides:1980nt_seesawII,Mohapatra:1980yp_seesawII,Schechter:1981cv_seesawII,Foot:1988aq_seesawIII} is very popular. 
In usual (natural) seesaw models light neutrino masses are $\sim m_D^2/M$ where the Dirac-type mass $m_D \sim 100$ GeV and $M$ is the Majorana mass of heavy
fermion which gets integrated out during the process. The mass of this heavy fermion, $M$, determines the scale of the seesaw models which needs 
to be very high ($\sim 10^{11}$ GeV) to avoid any fine tuning in $m_D$. As the natural scale of the seesaw is very high these models are suffere from a lack of testability. 
But it is also possible to construct low scale ($\sim$ TeV) models either importing some new fields \cite{Adhikari:2010yt} or incorporating higher-dimensional operators 
\cite{Babu:2009aq_HD,Bonnet:2009ej_HD,Kanemura:2010bq_HD,Bambhaniya:2013yca_HD,delAguila:2013yaa_HD}. 
These models not only generate the correct order of neutrino masses and mixing, but are also phenomenologically interesting as 
the scale of these theories are well within the reach of present experiments like the LHC.
These models are extended by some extra gauge symmetry and(or) new particles. The presence of these new fields 
might affect the evolution of the SM couplings, like gauge, Higgs quartic, and top Yukawa couplings if they couple to the SM 
particles. Hence it is necessary to examine the status of the SM vacuum once these new physics models come into play. Thus by using  knowledge
of the SM parameters and from the demand of vacuum stability\footnote{In this paper we are considering stability up to the Planck scale. We are not considering the metastability which does not require the vacuum to be bounded from below. If the decay life time of the vacuum is larger than the life time of the universe then that vacuum is metastable. But as our procedure concerns only boundedness of the scalar potential it fails to pin down the existence of the metastable vacuum.} the new parameters involved in the theory might be severely constrained. 
In the literature  the stability of the vacua was discussed in several 
scenarios considering beyond Standard Models (BSMs). These models are extended
by the extra gauge symmetry and (or) addition new particles. 
Quantum corrections of the quartic couplings depend on the spin of the particles belonging  to a particular model. The fermion loop contributions contain a relative minus `-' sign comapred to for the bosonic fields. Thus
the Yukawa couplings tend to spoil the stability unlike the gauge and other scalar self-couplings. 
Vacuum stability in different variants of see-saw models has been 
adjudged in Refs.~\cite{VS_seesaw_1,VS_seesaw_2,VS_seesaw_3,VS_seesaw_4,VS_seesaw_5,VS_seesaw_6,VS_seesaw_7} which has richer particle spectrum compared to 
the SM. In the context of gauge extensions, vacuum stability for the alternative left-right Symmetric Model has been discussed in Ref.~\cite{VS_ALRSM}.

In a theory involving multiple scalar fields the structure of the potential is complicated. The vacuum stability criteria depend 
on some combinations of the scalar quartic couplings. Moreover, the perturbativity (triviality) bounds also play crucial roles in finding a  consistent parameter space compatible with the choice of new physics scales. Non tachyonic scalar masses are guaranteed with these constraints.
It has been noted that some of the quartic couplings can be recast in terms of the heavy scalar masses and thus 
can be constrained from phenomenological point of view. On the contrary, few of them do not have that much impact 
on scalar masses rather they determine the splitting among the narrowly spaced massive scalar modes. 
Our present collider experiments still not sensitive to address that fine splittings thus those quartic couplings are beyond the reach of any experimental verification.
But those couplings can be constrained through vacuum stability, perturbativity (triviality) depending on the choice of scale of new physics.

In our study we have concentrated oLeft-Rightn the $U(1)_{B-L}$ extended models which are classified into two categories
:$SM\otimes U(1)_{B-L}$ 
or left-right (LR) symmetry. We have adopted two variants of the LR symmetric models containing (i) two $SU(2)$ triplet scalars  $\Delta_{L(R)}$, 
and (ii) two $SU(2)$ doublet scalars, $H_{L(R)}$. In section~\ref{sec:models} we introduce the basic structures of these models.
Then we include the renormalization group evolutions of all the necessary couplings and 
show how the vacuum stability, perturbativity (triviality) bounds constrain the parameter space of each models in section~\ref{sec:vac_stability}. 
We have analysed the structure of the potentials in detail and computed the criteria for vacuum stability using the formalism shown in
 reference~\cite{Arhrib:2011uy_vs}. All vacuum stability conditions corresponding to different models are listed in appendix~\ref
{app:Higgs_potential_VS}.

\section{Models}  \label{sec:models}
The Standard Model symmetry group is expressed as $SU(3)_C\otimes SU(2)_L \otimes U(1)_Y$. 
It has been noted in \cite{Marshak:1979fm_BL} that an extra $U(1)$ gauge symmetry 
along with the SM can provide solutions to some of the unaddressed issues in the Standard Model. 
These extra Abelian symmetry groups can, in general, originate from different high 
scale Grand Unified Theories (GUTs), like $SO(10), E(6)$. These larger groups contain $U(1)_{B-L}$ as a part of the intermediate gauge symmetries. 
In nonsupersymmetric GUT models the $U(1)_{B-L}$ breaking scale can be lowered as few TeV\footnote{This is also true for supersymmetric GUT models, see \cite{Chakrabortty:2009xm_jdGUT}.} \cite{Chakrabortty:2009xm_jdGUT}, 
which is consistent with unification pictures. In our present study we concentrate on TeV scale 
$U(1)_{B-L}$ extended models where neutrino mass generation can be explained. However, any high scale root of these models are not considered and kept for future work.

\subsection{$U(1)_{{B-L}}$}  \label{sec:model_BL}
The gauge group under consideration is $SU(3)_C\otimes SU(2)_L \otimes U(1)_Y \otimes U(1)_{B-L}$.
This minimal model contains an extra complex singlet scalar field $S$ and this extra $B-L$ symmetry is broken once it acquires 
vacuum expectation value ($vev$) \cite{Iso:2009nw_tevBL,Khalil:2010iu_tevBL,Basso:2010jm_U1BL}. 
Thus the $vev$  determines the symmetry-breaking scale of this symmetry 
and also the mass of the extra neutral gauge boson $Z_{B-L}$. For the purpose of our study we will focus only on
the relevant part of the Lagrangian, namely the scalar kinetic, and potential terms and the lepton Yukawa couplings. 
The scalar kinetic term is:
\be\label{eq:new-scalar_L}
 \mathscr{L}_s = (D^\mu \Phi)^{\dag} (D_\mu \Phi) + (D^\mu S)^{\dag} (D_\mu S) - V(\Phi,S). 
\ee
Here the potential $V(\Phi,S)$ is given as:
\be \label{eq:BL-potential}
V(\Phi,S )=m^2\Phi^{\dagger}\Phi + \mu ^2\mid S \mid ^2 + \lambda_1 (\Phi^{\dagger}\Phi)^2 +\lambda_2 \mid S \mid ^4 + 
\lambda_3\, \Phi^{\dagger}\Phi\mid S \mid ^2,  
\ee
where $\Phi$ and $S$ are the complex scalar doublet and singlet fields respectively.
After gauging away the extra modes and acquiring the $vev$s these fields are redefined as: 
\be \label{eq:min}
 \Phi  \equiv \left( \begin{array}{c} 0 \\ \frac{1}{\sqrt{2}}(v+\phi) \end{array} \right)\, , 
	\hspace{2cm}  S  \equiv \frac{1}{\sqrt{2}}(v_{_{B-L}}+s)\, ,
\ee 
where, EW symmetry breaking $vev$, $v$ and $B-L$ breaking $vev$, $v_{_{B-L}}$ are real and positive. 


We also find the scalar mass matrix in the following form:
\be \label{eq:mass-matrix}
\mathcal{M} = \left( 
		\begin{array}{lr}
                \lambda_1 v^2  & \frac{\lambda_3 v_{_{B-L}} v}{2}\\
                \frac{\lambda_3 v_{_{B-L}} v}{2} & \lambda_2 v_{_{B-L}}^2\\
               \end{array}
               \right)
               = \left( 
		\begin{array}{lr}
                \mathcal{M}_{11} \;&\; \mathcal{M}_{12}\\
                \mathcal{M}_{21} & \mathcal{M}_{22}\\
               \end{array}
               \right).
\ee
After diagonalising this mass matrix we construct two physical scalar states, a light $h$ and a heavy $H$, having masses $M_{h}$ and $M_{H}$, respectively, 
\be \label{eq:mass-eigenstates}
 M_{H,h}^2 = \frac{1}{2}\left[\mathcal{M}_{11}+\mathcal{M}_{22} \pm \sqrt{(\mathcal{M}_{11}-\mathcal{M}_{22})^2+4 \mathcal{M}_{12}^2}\right].
\ee
The scalar mixing angle, $\alpha$ can be expressed as:
\be \label{eq:alpha}
\tan(2\alpha) = \frac{2 \mathcal{M}_{12}}{\mathcal{M}_{11} - \mathcal{M}_{22}}=\frac{\lambda_3\,v\,v_{_{B-L}} }
{\lambda_1 v^2 - \lambda_2 v_{_{B-L}}^2}.
\ee
Using eqs.~\ref{eq:mass-eigenstates} and \ref{eq:alpha} the quartic coupling constants 
$\lambda_1,\,\lambda_2,\,$ and $\lambda_3$ can be recast in the following forms: 
\bea \label{eq:B-L_lambda}
 \lambda _1 &=& \frac{1}{4v^2}\Big\{\left(M_{H}^2+M_{h}^2\right)-\cos{2\alpha}\left(M_{H}^2-M_{h}^2\right) \Big\},\nonumber \\ 
 \lambda _2 &=& \frac{1}{4v_{_{B-L}}^2}\Big\{\left(M_{H}^2+M_{h}^2\right)+\cos{2\alpha}\left(M_{H}^2-M_{h}^2\right) \Big\},\nonumber \\ 
 \lambda _3 &=& \frac{1}{2\,v\,v_{_{B-L}}}\Big\{ \sin{2\alpha} \left({M_{H}^2-M_{h}^2} \right)\Big\}.
\eea
It can be noted from the last equation in eq.~\ref{eq:B-L_lambda} that we would get a duplicate set of solutions with inverted signs for  both $\alpha$ and $\lambda_3$. 
Hence one choice of positive $\alpha$ suffices as presented at section~\ref{sec:vac_BL}.


Due to the presence of an extra $U(1)_{B-L}$ gauge theory the SM gauge kinetic terms is modified by
\begin{equation}\label{La}
\mathscr{L}_{ {B-L}}^{KE} = -\frac{1}{4}F^{\prime\mu\nu}F^\prime _{\mu\nu}\, ,
\end{equation}
where,
\begin{eqnarray}\label{new-fs4}
F^\prime_{\mu\nu} &=& \partial _{\mu}B^\prime_{\nu} - \partial _{\nu}B^\prime_{\mu} \, .
\end{eqnarray}

The covariant derivative for $SU(2)_L\otimes U(1)_Y\otimes U(1)_{B-L}$ sector in this model is modified as
\begin{equation}\label{cov_der_B-L}
D_{\mu}\equiv \partial _{\mu} + ig_{2}T^aW_{\mu}^{\phantom{o}a} +ig_1YB_{\mu} +i(\widetilde{g}Y + g_{_{B-L}}Y_{B-L})B'_{\mu}\, .
\end{equation}
The SM gauge bosons $B_\mu$ and $W_\mu^3$ will mix with the new gauge boson $B'_\mu$ to create two massive physical fields $Z$ and $Z_{B-L}$ 
and one massless photon field $A$. Assuming there is no kinetic mixing at tree level, i.e., $\tilde{g}=0$ at the EW scale, 
the physical gauge-boson masses are given as
\bea\label{eq:B-L_MZ1}
M_Z^2 = \frac{1}{4}\left(g_{_1}^2+g_{_2}^2\right) v^2, \\ \label{eq:B-L_MZ2}
M_{Z_{B-L}}^2 = 4 g_{_{B-L}}^2 v_{_{B-L}}^2.
\eea

Along with the Standard Model particles, three right-handed neutrinos ($\nu_R$) are introduced\footnote{One right-handed neutrino ($Q_{B-L} = -1$) for each generation is required for the sake of gauge anomaly cancellation.}.
The relevant term of the Lagrangian of the Yukawa interactions can be written as
\be\label{eq:B-L_yukawa}
-\mathcal{L}_Y = y^l_{ij} \overline{l_{iL}}\, \widetilde{\Phi}\, \nu_{jR} + y^h_{ij}\; \overline{(\nu_R)_i^c}\,\nu_{jR} \,S + h.c.
\ee
where $\widetilde{\Phi} = i \sigma_2 \Phi^*$ with $\sigma_2$ being the  Pauli matrix. The second term of the above equation is the 
Majorana mass term. Note from the eq.~\ref{eq:B-L_yukawa} that conservation of $B-L$ charge requires thet  the singlet scalar field, $S$, must have $Q_{B-L} = -2 $. When the SM Higgs and singlet scalar $S$ acquire $vev$s the neutrino mass matrix takes the form 
\be
M_\nu = \left(\begin{array}{cc}  0  & m_D\\ m_D^T \;\;&\;\; m_R\\ \end{array} \right),
\ee
\noindent where $m_D = y^l \frac{v}{\sqrt{2}}$ and $m_R = \sqrt{2}\; y^h v_{_{B-L}}$. 
The light ($m_{\nu_l}$) and heavy ($m_{\nu_h}$) neutrino masses are 
\bea
m_{\nu_l} & = & -m_D^T m_R^{-1} m_D, \\ m_{\nu_h} & = & m_R. 
\eea

In this model heavy neutrino mass $m_R$ is also generated through the Yukawa terms unlike the gauge-invariant Majorana mass term in type-I seesaw.
It can be noted that with $m_R\sim \mathcal{O}$(TeV), $y^l$ needs to be very small to generate light neutrino masses $\sim \mathcal{O}$(eV).
But $y^h$ can be large $\sim \mathcal{O}$(1) as $v_{_{B-L}}$ is around TeV scale. Thus successful light neutrino mass generation 
does not constrain $y^h$. But as the heavy neutrino is also coupled to the SM-like Higgs, $y^h$ affects the vacuum stability of the 
scalar potential in this model and gets constrained. The gauge coupling $g_{_{B-L}}$, and, $vev$ of $B-L$ breaking scale are
also free parameters. In the following section we have shown how these parameters are constrained from vacuum stability  of the scalar potential and also from perturbativity (triviality) of the couplings.

\subsection{Left-Right Symmetry}  \label{sec:model_LRsymmetry}
The full LR symmetric gauge group is written as $SU(3)_C\otimes SU(2)_L\otimes SU(2)_R \otimes U(1)_{B-L}$.
The $SU(2)_R \otimes U(1)_{B-L}$ is broken to $U(1)_Y$ at a scale higher than the EW symmetry breaking one. 
Thus the hypercharge generator is a linear combination of $SU(2)_R$ and $U(1)_{B-L}$ generators. 
In this model, hypercharge, $Y$, can be reconstructed from the $SU(2)_R$ and $U(1)_{B-L}$ quantum numbers as: 
\be
Y = T_{3R}+(B-L)/2,
\ee
$T_{3R}$ being $3^{rd}$ component of $SU(2)_R$ isospin.

Here we briefly present  two variants of Minimal left-right Symmetric Models (MLRSMs): 
\begin{itemize}
 \item The scalar sector consists of a bidoublet ($\Phi$), one left-handed triplet ($\Delta_L$), and one right-handed triplet ($\Delta_R$)
 \cite{Pati:1974yy_LR,Mohapatra:1974hk_LR,Mohapatra:1974gc_LR,Senjanovic:1975rk_LR}. 
 \item Scalar sector consists of a bidoublet ($\Phi$), one left-handed doublet ($H_L$), and one right-handed doublet ($H_R$)
 \cite{Senjanovic:1978ev_LRD,Brahmachari:2003wv_LRD,Sarkar:2004hc_LRD}. 
\end{itemize}

\subsubsection{LR Model with Triplet Scalars}  \label{sec:model_LR_triplet}
The most generic scalar potential of this model with bidoublet and triplet scalars ($\Phi, \Delta_{L,R}$) is given in appendix~\ref{app:Higgs_potential_LRT}.
The explicit structures of the scalars can be presented in the following form
\begin{center}
 $
\hskip 25pt
\Phi =\left( \begin{array}{lr}
         \phi_1^0 \;&\; \phi_1^+\\ \\
         \phi_2^- & \phi_2^0
        \end{array} 
        \right)  \hskip 10pt , \hskip 10pt       
\Delta_{L,R}  = \left( \begin{array}{cc}
         \delta_{L,R}^+/\sqrt{2} & \delta_{L,R}^{++} \\  \\      
         \delta_{L,R}^0 & -\delta_{L,R}^+/\sqrt{2}
        \end{array} 
        \right) .
$
\end{center}
These fields transform under $SU(2)_L\otimes SU(2)_R \otimes U(1)_{B-L}$ gauge groups in the following manners: 
\be
\Phi \equiv (2,2,0), \hskip 19pt \Delta_R \equiv (1,3,2), \hskip 19pt \Delta_L \equiv (3,1,2).
\ee
Once neutral components of these scalars acquire vacuum expectation values, they can be written in the following form
\be\label{LRT_vev_struct} 
\left< \Phi \right>  =\left( \begin{array}{cc}
                        v_1 & 0\\
                        0      \;&\; v_2 e^{i\theta}
                       \end{array}\right) , \hskip 20 pt                       
\left< \Delta_L \right>  =\left( \begin{array}{lr}
                        0   & 0\\
                        v_L\; & \;0
                       \end{array}\right) , \hskip 20 pt                       
\left< \Delta_R \right>  = \left(\begin{array}{lr}
                        0   & 0\\
                        v_R \;&\; 0
                       \end{array} \right),
\ee
where, for simplicity we have chosen $v_2 = 0$ without loss of generality.
With these structures of the vacuum expectation values, symmetry breaking occurs in  two stages.
The symmetry group $SU(2)_L\otimes SU(2)_R \otimes U(1)_{B-L}$ breaks down to $SU(2)_L\otimes U(1)_Y$ by $v_R$
at high scale.  Consequently, the vacuum expectation value $v_1$ of bidoublet breaks $SU(2)_L\otimes U(1)_Y$ to $U(1)_{EM}$. 
So total number of Goldstone bosons will be six.  
Now the Higgs sector has 20 degrees of freedom (eight real field for the  bidoublet and six each for triplet fields). 
Hence, the remaining 14 fields will be massive scalars and they are as follows:
\begin{enumerate}
\item Two doubly charged scalars ($H^{\pm\,\pm}_1,H^{\pm\,\pm}_2$),
 \item Two singly charged scalars ($H^\pm_1,H^\pm_2$),
 \item Four neutral $CP-even$ scalars ($H_0^0,\;H_1^0,\;H_2^0,\;H_3^0\;$), 
 \item Two neutral $(CP-odd)$ pseudoscalars ($A_0^0,\;A_1^0\;$).
\end{enumerate}


Since already mentioned that the scale $v_R$ is much higher than the $vev$ of electroweak breaking  $v_1$, 
the scalar  masses can be expressed in leading-order terms\footnote{These leading order terms match exactly with the masses of the heavy 
scalars at scale $v_R$, i.e., before electroweak symmetry breaking (EWSB). After the EWSB, some correction terms are generated which are
proportional to the $v_1^2$. But as $v_R>>v_1$, the splitting among the masses of these heavy scalars are negligible compared to their 
relative masses. It is important to note that this `$\simeq$' will be replaced by `$=$' in eq.~\ref{eq:LR-Masses} when these masses are 
given at $v_R$ scale.} \cite{Duka:1999uc,Czakon:1999ue_WR_mudecay}
\bea \label{eq:LR-Masses}
M_{H_0^0}^2 &\simeq& 2 \, \la_1 \,v_1^2 ,\nonumber\\
M_{H_1^0}^2 &\simeq& \frac{1}{2} \la_{12}\, v_R^2, \nonumber\\
M_{H_2^0}^2 \simeq  M_{A_1^0}^2 \simeq M_{H_2^\pm}^2 &\simeq& 2 \, \la_5 \, v_R^2, \nonumber \\
M_{H_3^0}^2 \simeq  M_{A_2^0}^2 \simeq M_{H_1^\pm}^2 \simeq M_{H_1^{\pm\pm}}^2 &\simeq& \frac{1}{2} (\la_7 - 2\la_5) \, v_R^2, \nonumber \\
M_{H_2^{\pm\pm}}^2 &\simeq& 2\, \la_6 \, v_R^2.
\eea
$M_{H_0^0}$ is the Standard Model Higgs boson and denoted as $M_h$ from here onwards. For simplicity and to reduce the number of free parameters, we consider degenerate heavy scalars at the $v_R$ scale, i.e., 
$M_{H_1^0}=M_{H_2^0}=M_{H_3^0}=M_{H_{2}^{\pm\pm}} = M_H$. It is important to note that the remaining quartic 
couplings only contribute in the scalar masses as subleading terms and they are proportional to the $v_1^2$ at the 
ekectroweak symmetry-breaking scale (EWSB)  scale. Hence, $\la_2, \la_3, \la_4, \la_8, \la_9, \la_{10},$ and $\la_{11}$ 
induce only the relative mass splittings among these heavy scalars which are almost phenomenologically unaccessible at present experiments.


The kinetic term of scalar part  can be written as 
\be \label{eq:LRT_kinetic}
\mathcal{L}_{kin} = \Tr\Big[ (D_\mu \Phi)^\dagger (D^\mu\Phi) \Big] +\Tr\Big[ (D_\mu \Delta_L)^\dagger (D^\mu\Delta_L) \Big] 
                   +\Tr\Big[ (D_\mu \Delta_R)^\dagger (D^\mu\Delta_R) \Big], 
\ee
where,
\begin{eqnarray}\label{eq:LRT_cov_deriv}
D_\mu \Phi &=& \partial_\mu \Phi - i g_{_{2L}}\,T^a\, W_{L\mu}^a\,\, \Phi + i g_{_{2R}}\,\,\Phi \,\,T^a W_{R\mu}^a\,,  \\ 
D_\mu \Delta_{(L/R)} &=&\partial_\mu \Delta_{(L/R)}- i g_{_{(2L/2R)}}\Big[T^a W_{(L/R)\mu}^a \, , \, \Delta_{(L/R)}\Big]-ig_{_{B-L}}B_\mu\Delta_{(L/R)}\,.\nonumber
\end{eqnarray}

We choose the gauge couplings $g_{2L}$ and $g_{2R}$ for the $SU(2)_L$ and $SU(2)_R$  gauge groups 
respectively to be same for the sake of minimality of the model in terms of number of parameters.
After spontaneous breaking of  LR and EW symmetries,
two charged $W_{L/R}^{\pm}$ and two neutral $Z_{L/R}$ gauge bosons become massive, while photon $A$ remains massless:

\bea \label{eq:LRT_WZ_mass}
M_{W_L^\pm}^2 &=& \frac{1}{4}g_{_{2}}^2\,v_1^2 \;,\hskip 2cm
M_{W_R^\pm}^2 = \frac{1}{4}g_{_{2}}^2\,\left(v_1^2+2\,v_R^2\right),\\
M_{Z_{L,R}}^2 &=& \frac{1}{4} \Bigg[\left( g_{_{2}}^2v_1^2+2v_R^2(g_{_{2}}^2+g_{_{B-L}}^2)\right) \nonumber \\
              & & \mp  \sqrt{\left\{g_{_{2}}^2v_1^2+2v_R^2(g_{_{2}}^2+g_{_{B-L}}^2)\right\}^2-4g^2(g_{_{2}}^2+2g_{_{B-L}}^2)v_1^2v_R^2} \Bigg]. \nonumber
\eea


Under the gauge group $SU(2)_L\otimes SU(2)_R\otimes U(1)_{B-L}$ quarks and leptons are doublets,
\be L_{i(L/R)} = \left( \begin{array}{c}
         \nu_i\\ l_i
        \end{array} 
        \right)_{(L/R)}, \hskip 1.5cm
        Q_{i(L/R)} = \left( \begin{array}{c}
         u_i\\ d_i
        \end{array} 
        \right)_{(L/R)}.        
\ee
The most general lepton Yukawa Lagrangian can be written as, 
\be
-\mathcal{L}_Y = \bigg[\overline{L_{L}}\Big(y^l\, \Phi + \tilde{y}^l\,\tilde{\Phi}\Big) L_{R} + h.c\bigg] +
 y^h_L\, \overline{L_R^c}\, \widetilde\Delta_L \, L_L +y^h_R\, \overline{L_L^c}\, \widetilde\Delta_R \, L_R, 
\ee
here, $\tilde\Phi = i\sigma_2 \Phi^* $ and $\widetilde\Delta_{L/R} = i \sigma_2 \Delta_{L/R}$. Here we have considered that the Yukawa matrices are
diagonal\footnote{There exist two different discrete symmetries which can relate Left and Right handed fields\cite{LR_at_LHC}. Yukawa matrices
are diagonal as we have considered the parity operation as defined in \cite{Duka:1999uc} to relate $L$ and $R$ fields.}. 
The neutral fermion masses are generated once the $\Phi$ and $\Delta$ acquire $vev$.
The neutral fermion mass matrix is given as
\be
M_\nu = \left( \begin{array}{lr}
         m_\nu^{II} \;&\; m_D\\
         m_D^T & m_R
        \end{array}
	  \right), 
        \hskip 0.5cm     
m_D =   \frac{1}{\sqrt{2}}y^l v_1,\;\; m_R = \sqrt{2} y^h v_R,\;\;m_\nu^{II}=\sqrt{2} y^h v_L,  
\ee
here, $y^h_L=y^h_R=y^h$ because of left-right symmetry. Thus the light neutrino mass
\be
m_{\nu_l}=m_\nu^{II}-m_D^T m_R^{-1} m_D,
\ee
is generated through type-II (first term) and type-I (second term) seesaw mechanisms.

As the $vev$ of the left-handed triplet scalar is constrained from $\rho$ parameter of the SM it cannot be larger than $\sim \mathcal{O}$(few GeV).
Thus it is indeed possible to generate light neutrino masses $\sim$ eV with $v_L \sim$ eV while the neutrino Yukawa coupling can be 
$\sim \mathcal{O}(1)$. In our further analysis we consider $v_L$=0, thus type-II seesaw is absent here. 
The heavy neutrino mass $m_R$ is also generated through the Yukawa terms and proportional to $v_R$.
It can be noted that with $m_R \sim \mathcal{O}$(TeV), the Dirac term $m_D$ needs to be 
very small to generate light neutrino masses $\sim \mathcal{O}$(eV).
But $y^h$ can be as large as $\sim \mathcal{O}$(1) even when $v_{R}$ is around TeV scale. Thus successful light neutrino mass generation 
is still possible keeping $y^h$ as large as $\sim \mathcal{O}(1)$. But $y^h$ affects the vacuum stability of 
the scalar potential in this model as the heavy neutrino is 
also coupled to the SM like Higgs. In the following section we have shown how these parameters are 
constrained due to vacuum stability and perturbativity (triviality).

It has been noted that the minimal left-right symmetric model is constrained by  flavour-changing neutral currents (FCNCs) 
\cite{Deshpande:1990ip_LRT,FCNC1,FCNC2,FCNC3}. The model we have worked with contains the bidoublet whose one of the $vev$ is zero.
Thus there is no FCNC problem in this model. There are also constraints from neutral kaon mixing, i.e., the kaon mass difference. Our 
choice of $v_R$ scale and the masses for the heavy neutral scalars takes care of those bounds. As the $vev$s and the Yukawa couplings 
in our scenario are real there is neither a source of nor spontaneous or explicit CP-violation. But since we have considered the Yukawa matrices
to be diagonal we will boil down to the trivial, i.e., identity CKM and PMNS matrices. To fit all the masses and mixings we need to go for
the non-minimal extension of this model and that certainly modify the set of RGEs that we have used here.

\subsubsection{LR Model with Doublet Scalars}  \label{sec:model_LR_doublet}
In this case the scalar sector consists of a bidoublet ($\Phi$), 
one left-handed  doublet ($H_L$), and one right-handed doublet ($H_R$). 
The scalar potential is depicted in appendix~\ref{app:Higgs_potential_LRD}.
In terms of $SU(2)_L \otimes SU(2)_R \otimes U(1)_{B-L}$ gauge group these fields can be written as,
\be
\Phi \equiv (2,2,0),\hskip 20pt H_L \equiv (2,1,1), \hskip 20pt \textrm{and},\hskip 10pt H_R \equiv (1,2,1).
\ee
\noindent The structure of $H_{L/R}$ is written as,\\
\be
H_{L/R} = \left( \begin{array}{c}
                        h_{L/R}^0 \\ \\
                         h_{L/R}^+
                       \end{array}\right) .
\ee
The neutral components of $\Phi$ and $H_{L/R}$ acquire the vacuum expectation values:
\be
\left< \Phi \right>  =\left( \begin{array}{cc}
                        v_1 & 0\\
                        0      & v_2 e^{i\theta}
                       \end{array}\right) , \hskip 20 pt                       
\left< H_L \right>  =\left( \begin{array}{c}
                        0  \\
                        v_L
                       \end{array}\right) , \hskip 20 pt                       
\left< H_R \right>  = \left(\begin{array}{c}
                        0   \\
                        v_R 
                       \end{array} \right).
\ee
As before, we put $v_2  = 0$. 
The scalar sector consists of sixteen real scalar fields out of which six will be Goldstone bosons. Finally we will have four CP-even scalars
 and two CP-odd scalars and two charged scalars. Among the CP-even scalars one is Standard Model Higgs boson with mass $M_h$ and 
 other three are taken as degenerate heavy scalars having mass $M_H$.
The parameters in the Higgs potential can be recast in terms of the masses of the neutral and charged scalars. 
The details about the scalar sector have been discussed in Ref.~\cite{Holthausen:2009uc_LRD}.
The gauge sector is similar to the previous case, i.e. the  LR model with triplet scalars.

In the limit $v_R >> v_1$ and assuming all the heavy scalars are degenerate, we have 
\be
f_1 =  (M_{H}/v_R)^2 = \kappa_1 = - \kappa_2, 
\ee
whereas, minimisation of the  potential requires: 
\be
\frac{v_1^2}{v_R^2}  =  \frac{f_1-2\beta_1}{4\la_1}. \nonumber
\ee

The structure of the covariant derivative in this model is very similar to that for the triplet scenario, see eq.~\ref{eq:LRT_cov_deriv}
\begin{eqnarray}\label{eq:LRD_cov_deriv}
D_\mu \Phi &=& \partial_\mu - i g_{_{2L}}\,T^a\, W_{L\mu}^a\,\, \Phi + i g_{_{2R}}\,\,\Phi \,\,T^a W_{R\mu}^a,  \\ 
D_\mu H_{(L/R)} &=&\partial_\mu H_{(L/R)}- i g_{_{(2L/2R)}} T^a W_{(L/R)\mu}^a\,H_{(L/R)}-ig_{_{B-L}}B_\mu H_{(L/R)}.\nonumber
\end{eqnarray}
Following the previous convention we also set $g_{_{2L}} = g_{_{2R}} = g_{_{2}} $.
After spontaneous breaking of $SU(2)_L\otimes SU(2)_R\otimes U(1)_{B-L}$ symmetry, 
two charged $W_{L/R}^{\pm}$ and two neutral $Z_{L/R}$ gauge bosons become massive, while photon $A$ remains massless 

\bea \label{eq:LRD_WZ_mass}
M_{W_L^\pm}^2 &=& \frac{1}{4}g_{_{2}}^2\,v_1^2 \;,\hskip 2cm
M_{W_R^\pm}^2 = \frac{1}{4}g_{_{2}}^2\,\left(v_1^2\,+\,v_R^2\right),\\
M_{Z_{L,R}}^2 &=& \frac{1}{8} \left[\left( 2g_{_{2}}^2v_1^2+v_R^2(g_{_{2}}^2+g_{_{B-L}}^2)\right)
                \mp\sqrt{4g_{_{2}}^4\,v_1^4+(g_{_{2}}^2+g_{_{B-L}}^2)v_R^4-4g_{_{2}}^2 g_{_{B-L}}^2v_1^2 v_R^2} \right]. \nonumber
\eea

In left-right symmetric model with doublet scalar leptonic part of the Yukawa interaction can be written as 
\be
-\mathcal{L} =  \bar{L}_L \Big( y_1 \Phi + y_2 \tilde{\Phi} \Big) L_R + h.c.
\ee
where $SU(2)_L\otimes SU(2)_R$ quantum numbers of $L_L$ and $L_R$ are (2,1) and (1,2) 
respectively. So from this Lagrangian the Dirac mass term for the neutrinos can be written as 
\be
m_D = y_1 v_1.
\ee

Here, it is not possible to write the renormalizable Majorana mass term for the light and heavy neutrinos. But we can add non-renormalizable effective terms as
\be \label{eq:effective_term}
\mathcal{L}_{eff} = \frac{\eta_L}{M} L_L L_L H_L H_L + \frac{\eta_R}{M} L_R L_R H_R H_R,
\ee
where, $M$ is some very high scale and $\eta$'s are dimensionless parameters denote the strength of these non-renormalizable couplings. 
Once $H_R$ acquires the $vev$ the right-handed neutrino mass is generated as
$$
m_{R} \simeq \frac{\eta_R v_R^2}{M}.
$$
Here we consider that $\langle H_L\rangle=v_L$ = 0, thus this effective term does not contribute to the light neutrino mass. 
The neutrino mass matrix in ($\nu_l$, $\nu_h$) basis reads as
\be
M_\nu =  \left( 
		\begin{array}{lr}
                0  & m_D\\
                m_D^T \;&\; m_{R}\\
               \end{array}
               \right),
\ee
and the light neutrino mass can be written as 
\be
m_{\nu_l} = - m_D^T \; m_{R}^{-1} \; m_D,
\ee
which is a variant of the type-I seesaw mechanism. 

In the left-right symmetric model associated with two doublet scalars, neutrino masses cannot be generated 
through type-II seesaw mechanism due to the lack of left-handed triplet scalar\footnote{Although, 
through an effective operator the Majorana mass term for light neutrino can be generated, see eq.~\ref{eq:effective_term}. 
But this contribution is absent here as we have set $v_L=0$.}. 
Thus the type-I seesaw mechanism is the natural choice in this case. 
But the right-handed neutrino masses are generated through an effective operator suppressed by a heavy scale. This may provide a possible explanation how the right-handed 
neutrinos can be lowered to TeV scale. Here, the correct order of light neutrino masses are generated if the Dirac-type neutrino Yukawa coupling needs to be very small unless one considers the special textures for the Dirac Yukawa couplings. 
Then vacuum stability is automatically satisfied as these Dirac Yukawa couplings are much smaller. 
Thus here only the quartic couplings get constrained through the vacuum stability, 
perturbativity (triviality) of the couplings. Within a framework very similar to this it is indeed possible to generate light neutrino masses of correct order 
without lowering the Yukawa coupling as the light neutrino masses are independent of $v_R$ but suppressed by some high scale
\cite{Malinsky:2005bi_LRDnu,Chakrabortty:2010zk_LRDnu}. 
On that case the vacuum stability constraints cannot be avoided and play the most crucial role in constraining the Yukawa couplings and other parameters.

\section{Vacuum Stability}  \label{sec:vac_stability}

The presence of new physics introduces exotic non-SM particles in the theory and 
if they couple to the SM fields then the renormalization group evolutions (RGEs) of the 
Higgs quartic coupling ($\la_h$) will be modified. Moreover, additional quartic interactions of extra scalar fields should also be introduced. 
Extended gauge interactions from the larger gauge groups as well as Yukawa interactions would contribute to these evolution equations.
Now the question arises of whether or not the vacuum is stable in the  presence of the new physics. 
In particular, when we have narrowed down a preferred range of the Higgs mass between 123-127 GeV, the new physics could be constrained by the vacuum stability criteria. 
To adjudge the stability of these models we have considered the one loop RGEs of all the required parameters. In passing we would like to mention that 
the allowed parameter space in our analysis is the minimal set which will be extended once one includes the higher order renormalization group (RG) effects.
The RGEs for SM and each of the $B-L$ models which are used in our calculation are given in appendix. 
Since we are dealing with the TeV scale models, all the SM RGEs will be 
modified once the new physics effects are switched on. Thus from EW scale to TeV (specific values are dictated in plots) the RGEs 
will be SM like and from the TeV scale to the Planck scale they will be the modified ones, and during the process 
proper matching conditions are incorporated at the TeV scale.

\subsection{$U(1)_{B-L}$ Model}   \label{sec:vac_BL}

It is clear from the structure of the potential as shown in eq.~\ref{eq:BL-potential} for the  $U(1)_{B-L}$ model, 
that the vacuum stability conditions are different from that for the SM due to the  presence of extra singlet scalar. 
If all the quartic couplings are positive, the potential will be trivially bounded from below, i.e., vacuum is stable and these stability conditions read simply as $\lambda_{1,2,3}>0$. But it is indeed possible to allow $\lambda_3$ to be negative and still have the vacuum be stable. Thus vacuum stability conditions beyond the trivial ones allow larger parameter space and need to be accommodated in these conditions.
We find the non-trivial vacuum stability criteria using the proposal dictated in \cite{Arhrib:2011uy_vs} and shown in appendix~\ref{app:vac_B-L},
\bea 
4 \lambda _1 \lambda _2 - \lambda _3^2>0, \,\nonumber \\
\lambda _1 > 0,\;\;\;\; \lambda _2 > 0. \, 
\label{eq:U1B-L_vs}
\eea
Together with these we have also incorporated perturbativity constraints on quartic couplings by demanding upper limit, i.e., $|\lambda_i| < 1 \;(i=1,2,3)$.

Noting down from eqs.~\ref{eq:mass-eigenstates} and \ref{eq:alpha} that 
the physical Higgs field is an admixture of two scalar fields $\phi$ and $s$, in our study the scalar mixing angle $\alpha$ is considered 
to be a free parameter instead of the quartic couplings $\lambda_i (i=1,2,3)$. 
This model consists of two different scales in the theory, those are  EW scale and $B-L$ symmetry breaking scale. 
Thus two RGEs are invoked for the analysis. As we have two Abelian couplings in this model, there might be
mixing between them \cite{Holdom:1985ag_2U1,delAguila:1988jz_2U1}. 
To simplify the situation, and off course without hampering any other conclusions, 
we impose no mixing between the  $Z_{B-L}$ and $Z$ gauge bosons at the tree-level. 
This is followed from the condition $\widetilde{g}(Q_{EW}) = 0$ as already discussed above eq.~\ref{eq:B-L_MZ1}. 
As a consequence $B-L$ breaking $vev$ $v_{_{B-L}}$ relates to the new $Z_{B-L}$ boson mass given as in eq.~\ref{eq:B-L_MZ2}. 
For demonstration, we have picked the perturbative value of this additional gauge coupling at breaking scale as, $g_{_{B-L}}=0.1$. 
For simplicity we further assume heavy neutrinos are degenerate and fixed at $m^{1,2,3}_{\nu_h} \equiv m_{\nu_h} \simeq 200$ GeV, which are 
within the allowed values.
We have used central value of light Higgs mass ($M_{h}$) at 125 GeV, top quark mass at 173.2 GeV and strong coupling constant $\alpha_s$ at 0.1184. Thus remaining free parameters in our study are 
$M_{H}$, $\alpha$ and $v_{_{B-L}}$. We have explored the correlated constraints on these parameters from vacuum stability.


\begin{figure}[tb]
\begin{center}
\includegraphics[width=7cm,angle=-90]{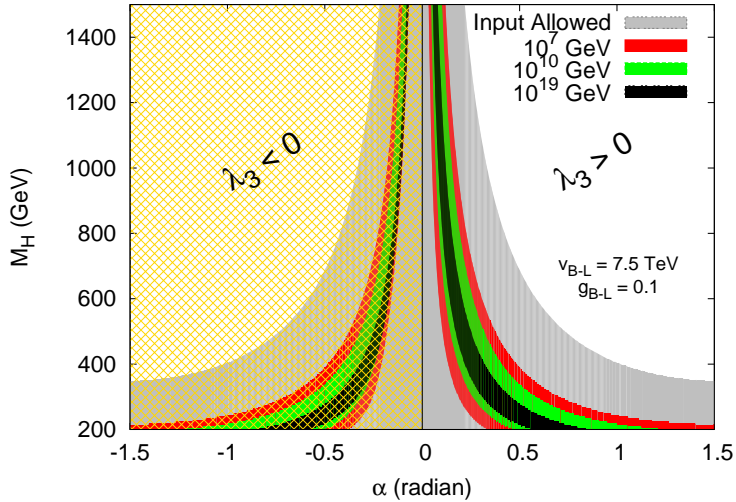}
\caption{The allowed parameter space in heavy Higgs mass ($M_H$) and scalar mixing angle $(\alpha)$ 
 plane, consistent with vacuum stability and perturbativity bounds are shown. The grey region is the domain of allowed input parameters. The red, green, and black sub-parameter spaces show the domain of $M_H$ and $\alpha$ for which this $B-L$ theory is valid till $10^7$, $10^{10}$
 and $10^{19}$ GeV respectively. The Majorana neutrino mass is fixed at 200 GeV and $B-L$ breaking
 $vev$ ($v_{_{B-L}}$) is set at 7.5 TeV. The $U(1)_{B-L}$ gauge coupling is taken to be 0.1 which implies $M_{Z_{B-L}}$=1.5 TeV. 
The shaded region satisfy $\lambda_3<0$ (as well as $\alpha < 0$ from eq.~\ref{eq:B-L_lambda}).  Thus the non-trivial vacuum stability conditions are being satisfied in this region. These conditions are stringent than the trivial one that applied in the positive $\alpha$ region. Although the pattern of the allowed parameter space is very similar for both positive and negative $\alpha$ region, the $\alpha>0$ region covers larger parameter space.
}
\label{fig:B-L_1}
\end{center}
\end{figure}

The set of RGEs of different couplings that we have used in our analysis are encoded in appendix~\ref{app:RG_BL} \cite{Basso:2010jm_U1BL}. 
The parameter space consistent with vacuum stability in heavy Higgs mass ($M_H$) and scalar mixing angle $(\alpha)$ 
 plane is depicted in figure~\ref{fig:B-L_1}. All the couplings are perturbative through out their evolutions. The grey region is the domain of allowed input parameters. 
 The red, green, and black sub-parameter spaces show the domain of $M_H$ and $\alpha$ for which this $B-L$ theory is valid till $10^7$, $10^{10}$
 and $10^{19}$ GeV respectively. In this figure, for a particular heavy scalar mass each of this allowed domain is restricted at some  minimum (maximum) value of $\alpha$ due to the vacuum stability (perturbativity) of the quartic couplings.
 The  Majorana neutrino mass is fixed at 200 GeV and $B-L$ breaking
 $vev$ ($v_{_{B-L}}$) is set at 7.5 TeV. The $U(1)_{B-L}$ gauge coupling is taken to be 0.1 which implies $M_{Z_{B-L}}$=1.5 TeV consistent  with present experimental bounds \cite{LHC_Zprime}. 
 The yellow shaded region  posses the set of allowed parameters for $\lambda_3<0$ (as well as $\alpha < 0$ from eq.~\ref{eq:B-L_lambda}).
Though the pattern of the allowed parameter space in positive $\lambda_3$ region is very similar, it is not exactly symmetric. 
The outer boundaries above of each color in figure~\ref{fig:B-L_1} matches exactly for both the positive- and negative- $\alpha$ region. 
This is not surprising because outer boundary is determined by the perturbativity of the couplings and thus not affected by the vacuum stability conditions which are different for different signs of $\lambda_3$. 
However, the lower boundaries are outcome of the demand to satisfy the criteria of vacuum stability.
Allowed parameters in the yellow shaded region (which represents $\lambda_3<0$) in  figure~\ref{fig:B-L_1} 
are reflected by the non-trivial vacuum stability condition in eq.~\ref{eq:U1B-L_vs}, 
which sequentially plays a role in determining the lower boundaries in the allowed parameters.
Thus expectedly in the positive $\alpha$ region the allowed parameter space is larger than that for negative $\alpha$.
Also, note that $\alpha=0$ leads to the decoupling limit when the heavy scalar will not affect the vacuum stability. The parameter space has also shrunk as the validity of the model must be closer to the Planck scale as can be inferred from the figure~\ref{fig:B-L_1}.

\begin{figure}[tb]
\begin{center}
\includegraphics[width=5.4cm,angle=-90]{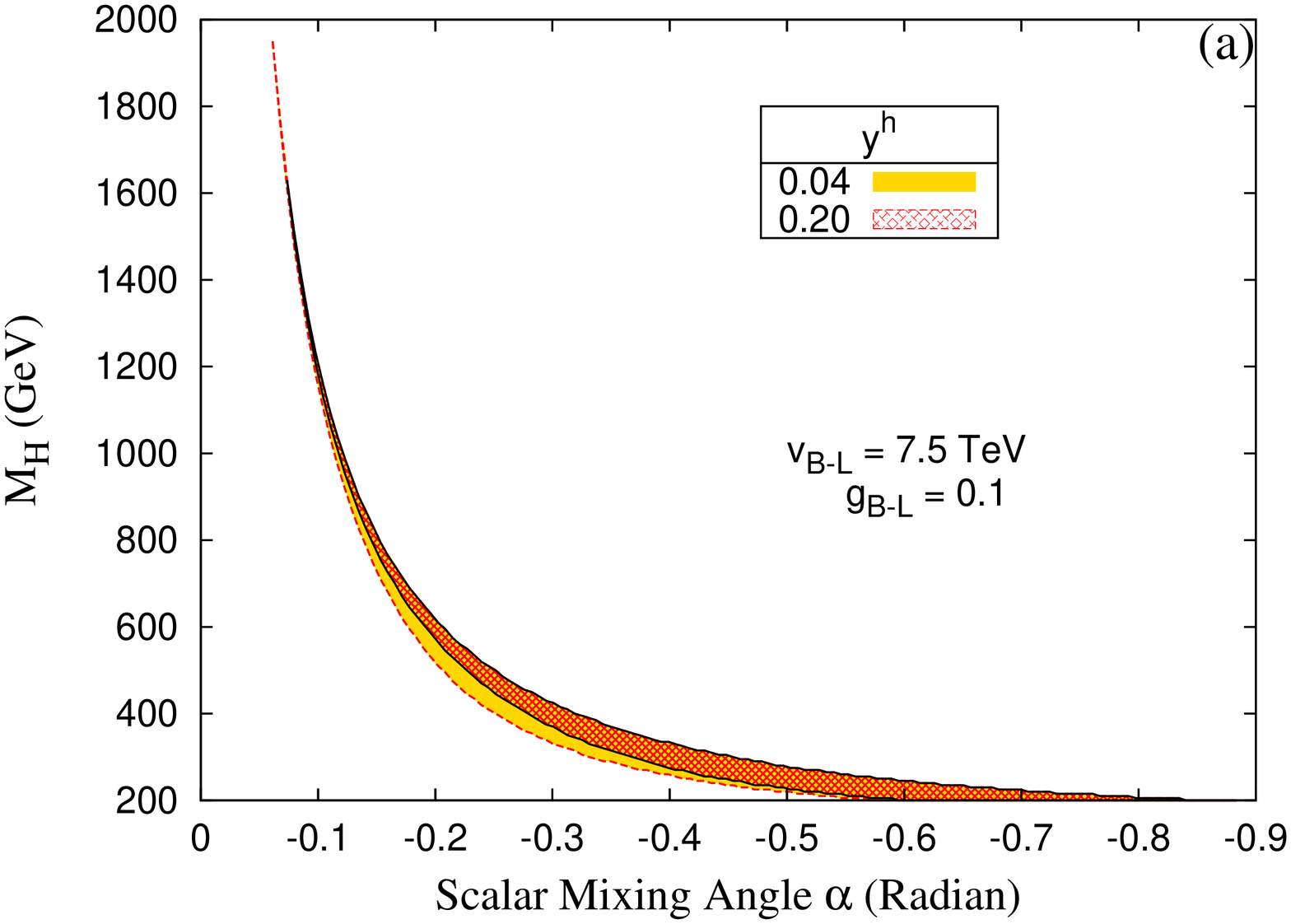}
\includegraphics[width=5.4cm,angle=-90]{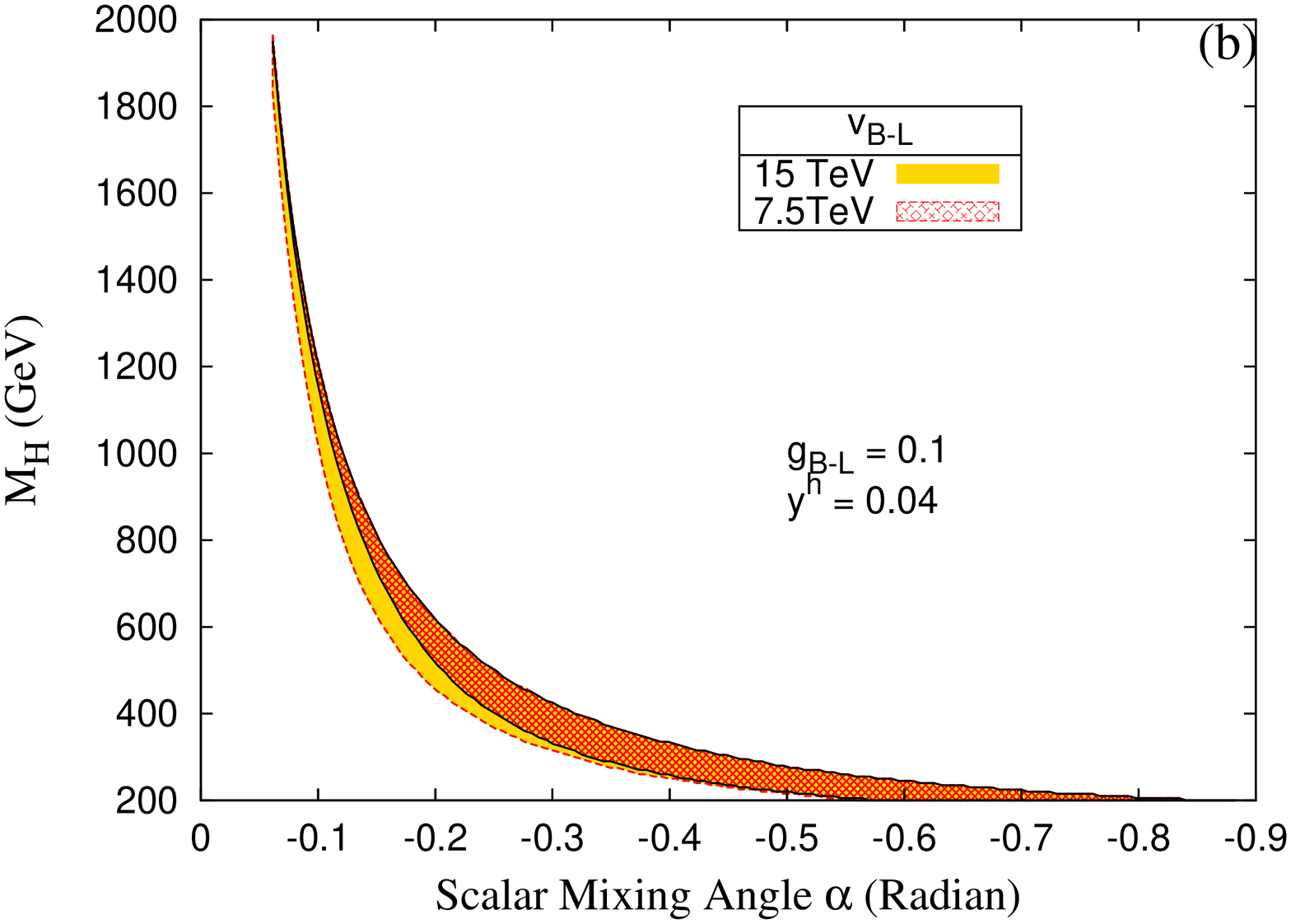}
\includegraphics[width=5.4cm,angle=-90]{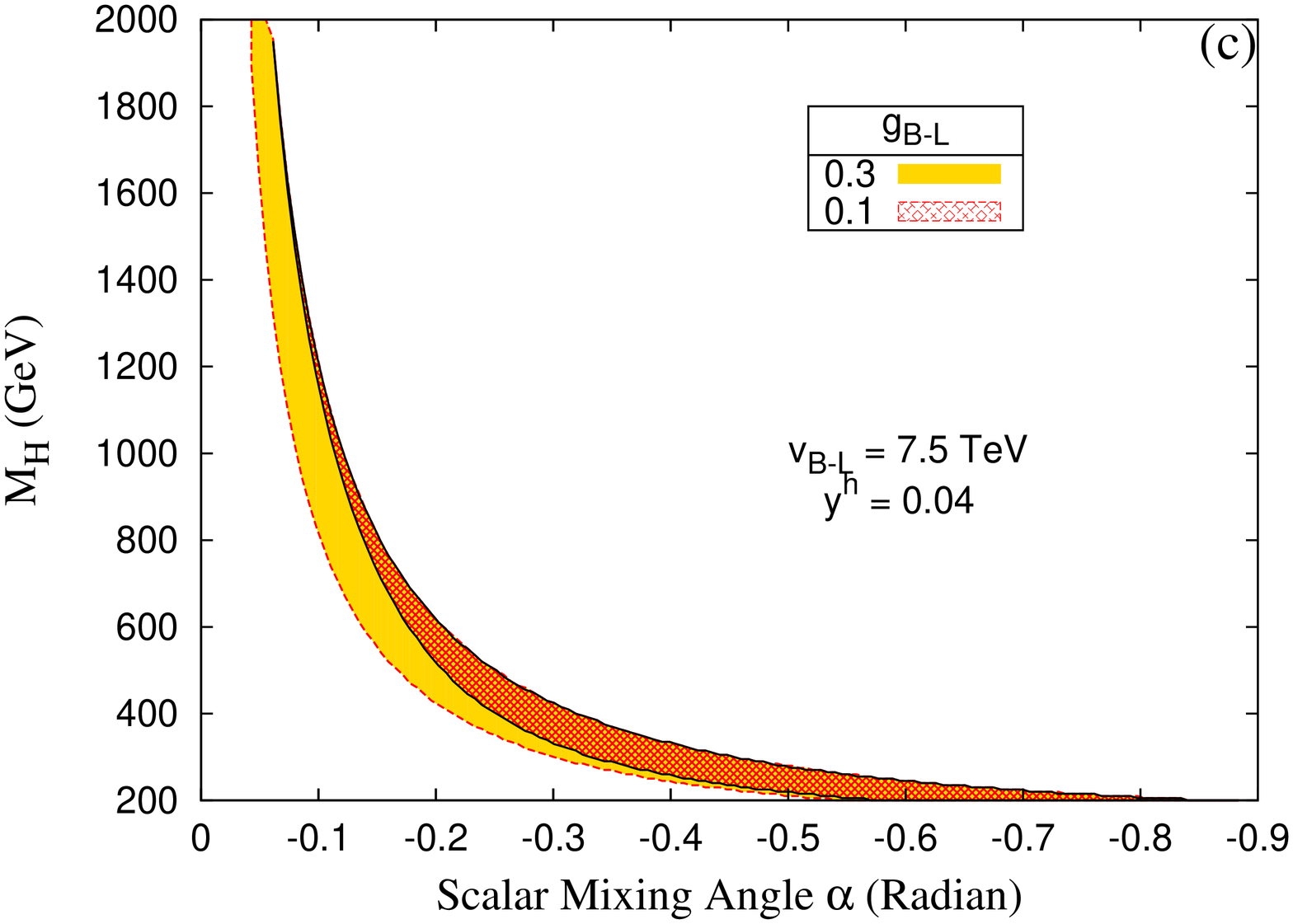}
\caption{Allowed parameter space in $M_H-\alpha$ plane, with $\alpha$ varying between $[0,-\pi/2]$, consistent with vacuum stability and 
perturbativity (triviality) bounds up to the Planck scale.
Figure (a): The Majorana neutrino Yukawa coupling $y^h$ is varied keeping $v_{_{B-L}}$ and $g_{_{B-L}}$ fixed. 
Figure (b): Two different set of $B-L$ breaking $vev$, $v_{_{B-L}}$ are chosen keeping $g_{_{B-L}}$ and $y^h$ fixed.
Figure (c): In this plot $g_{_{B-L}}$ varies where $v_{_{B-L}}$ and $y^h$ are kept constant. 
In our analysis any value of $g_{_{B-L}}$ for $v_{_{B-L}}= 7.5$ TeV more than 0.34 are disallowed as the 
coupling becomes non-perturbative before Planck scale. 
Corresponding regions for positive $\alpha$ are not shown here, as they remain unaffected and are the same as those  
given in blue strip in figure~\ref{fig:B-L_1} owing to the trivial conditions.}
 \label{fig:B-L_2}
 \end{center}
 \end{figure}

 To study the dependence of different parameters as shown in figure~\ref{fig:B-L_1}, we plot the
allowed parameter space in $M_H-\alpha$ plane which remains consistent with vacuum stability 
and where all the couplings are perturbative till the Planck Scale.
In figure~\ref{fig:B-L_2}(a) Majorana neutrino Yukawa coupling $y^h$ is varied keeping $v_{_{B-L}}$ and $g_{_{B-L}}$ fixed. As the $y^h$ increases, the allowed parameter space is shrunk since the Yukawa coupling affects the quartic couplings negatively in their RG evolutions. 
Thus larger Yukawa couplings spoil the vacuum stability. 
In figure~\ref{fig:B-L_2}(b) shows the dependence on $B-L$ breaking $vev$ for fixed $g_{_{B-L}}$ and $y^h$.  $v_{_{B-L}}$ determines the scale of new physics 
beyond the Standard Model, i.e., from where the RGEs are being modified
due to the presence of new particles. The larger $v_{_{B-L}}$ implies that new set of RGEs come to play later. In $B-L$ extended model 
$\lambda_3$ is inversely proportional to $v_{_{B-L}}$ at EW scale (see eq.~\ref{eq:B-L_lambda}). Thus for same set of values of $M_H$ and $\alpha$, $\lambda_3$ is smaller for larger $v_{_{B-L}}$ at 15 TeV. The RGE of $\lambda_3$ is such that for our choice of parameters it grows with mass scale. Thus 
there is a possibility of generating large $\lambda_3$ such that vacuum stability and perturbativity conditions are not validated at 
some higher scale. This plot therefore shows that it is possible to have larger allowed 
parameter space for larger $v_{_{B-L}}$.
Finally in figure~\ref{fig:B-L_2}(c), $g_{_{B-L}}$ varies where $v_{_{B-L}}$ and $Y_{\nu}$ are kept constant. 
As the larger values of the gauge couplings affect
the RGEs of the quartic couplings positively, the vacuum stability is improved. Thus with the larger value of gauge coupling 
the larger parameter space is allowed. But the $U(1)$ couplings increases with the mass scale. Hence the couplings with much larger values at 
low scale might be non-perturbative in the high scale. In our analysis, when $v_{_{B-L}}$ is at 7.5 TeV, any value of $g_{_{B-L}}$ 
more than 0.34 are disallowed as the coupling becomes non-perturbative before Planck scale.

\subsection{left-right Symmetry} \label{sec:vac_LR}

\subsubsection{LR Model with Triplet Scalars} \label{sec:vac_LR_triplet}

In this model the scalar potential for the left-right Symmetric model with triplet scalar as shown in the
appendix~\ref{app:Higgs_potential_LRT} contains many quartic couplings.
To find the condition of vacuum stability we have considered all two-fields, three-fields and four-fields directions 
and find their stability criteria. Detailed field directions corresponding to the potential together with calculated
stability conditions are listed in appendix~\ref{app:vac_LR_triplet}. 
Finally, the effective non-trivial vacuum stability conditions which are necessary and sufficient are
\bea
\lambda_1 > 0 ,\hskip 15pt \lambda_5 > 0 ,\hskip 15pt \la_5+\lambda_6 >0, &  & \nonumber \\
\la_5+2\,\lambda_6 >0, \hskip 10pt \hskip 10pt \lambda_{12} -2 \, \sqrt{\lambda_1 \lambda_5} < 0. & & 
\label{eq:LR_vs}
\eea
Along with the above conditions, we find an additional condition $\lambda_{12}>0$ from eq.~\ref{eq:LR-Masses}.

\begin{figure}[tb]
\begin{center}
\includegraphics[width=7cm,angle=-90]{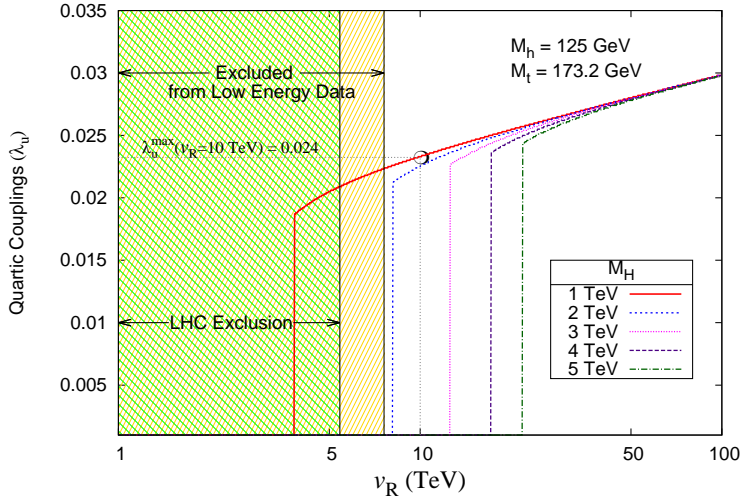}
 \caption{Constraints on universal quartic coupling $\la_u$ ($\equiv$ $\la_2$, $\la_3$, $\la_4$, $\la_8$, $\la_9$, $\la_{10}$, $\la_{11}$) for 
 LR model with triplet scalars in low $v_R$ region. Yellow Shaded region is disallowed from low energy data ($M_{W_R}>3.5$ TeV) and 
 green shaded region is excluded from direct search at LHC ($M_{W_{R}}>2.5$ TeV).}
 \label{fig:LRT-vR-l-uinversal}
 \end{center}
 \end{figure}
 
The renormalization group evolutions that we have considered in our analysis are depicted in appendix~\ref{app:RG_LR_triplet} \cite{Deshpande:1990ip_LRT}. 
In figure~\ref{fig:LRT-vR-l-uinversal} we show the constraints on universal quartic coupling 
$\la_u$ ($\equiv$ $\la_2$, $\la_3$, $\la_4$, $\la_8$, $\la_9$, $\la_{10}$, $\la_{11}$) for 
LR model with triplet scalars in low $v_R$ region. Yellow Shaded region is disallowed from low energy data ($M_{W_R}>3.5$ TeV) 
\cite{Beall:1981ze_kLkS,Langacker:1989xa_WR,Czakon:2002wm_mudecay,Chakrabortty:2012pp_WR_mudecay} and 
 green shaded region is excluded from direct search at LHC ($M_{W_{R}}>2.5$ TeV) \cite{Nemevsek:2011hz_LHCWR,Ferrari:2000sp_LHCWR,Chatrchyan:2012oaa_LHCWR,Aad:2012hf_LHCWR}.
 These limits can be extracted using the eq.~\ref{eq:LRD_WZ_mass}. 
In our analysis we also set Majorana Yukawa, $y^h$ at 0.25.
 We note that, for any particular heavy scalar mass ($M_H$), universal quartic coupling $\la_u$ is disallowed above 
 the corresponding line shown in the figure. For example, as seen from the plot, maximum allowed value of the universal quartic coupling is  0.024
 if one consider LR breaking scale at 10 TeV and heavy scalar mass at 1 TeV. Allowed maximum
quartic coupling is lowered for heavier scalar which can be understood from vacuum stability and perturbativity.

\begin{figure}[tb]
\begin{center}
\includegraphics[width=5cm,angle=-90]{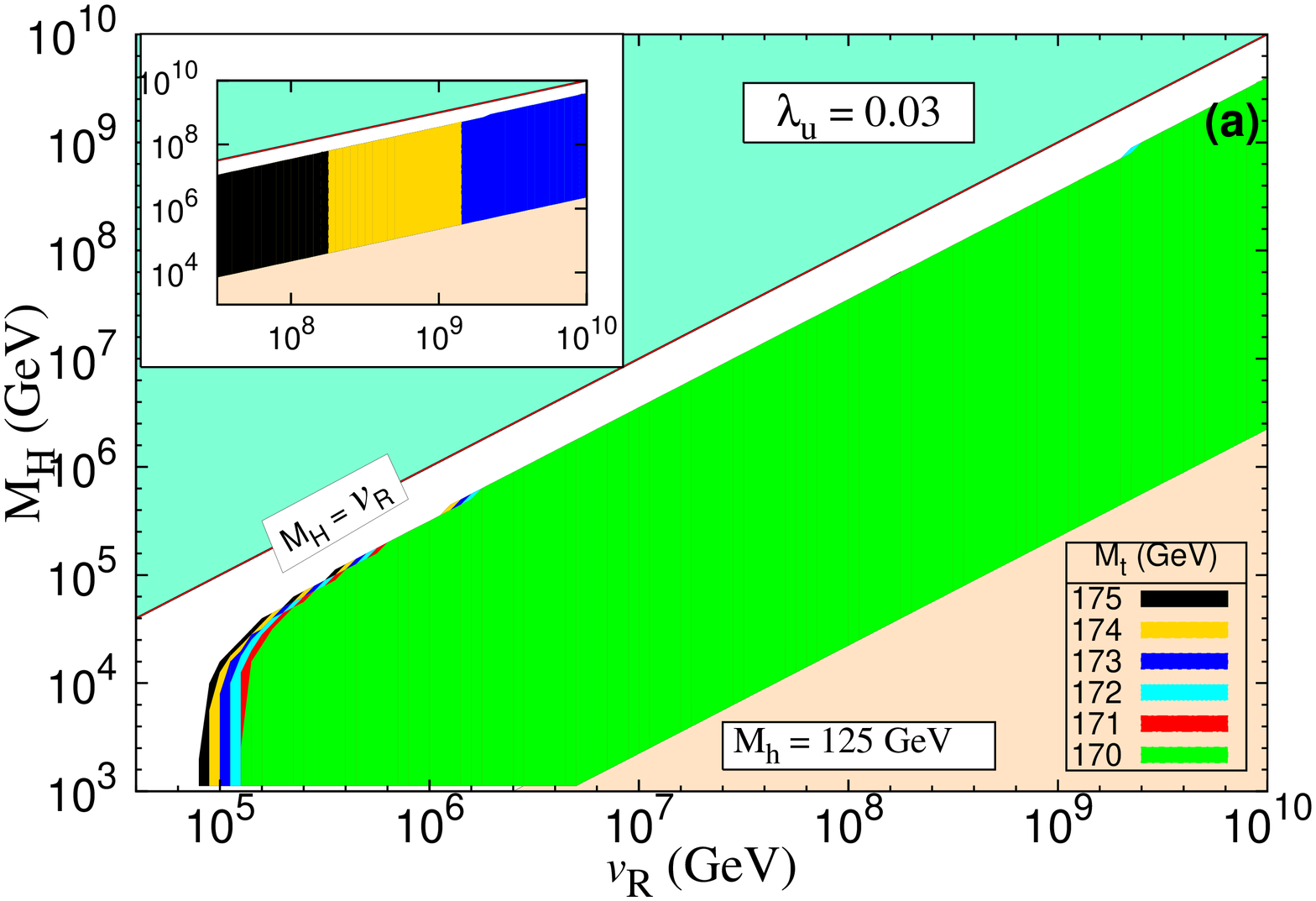}
\includegraphics[width=5cm,angle=-90]{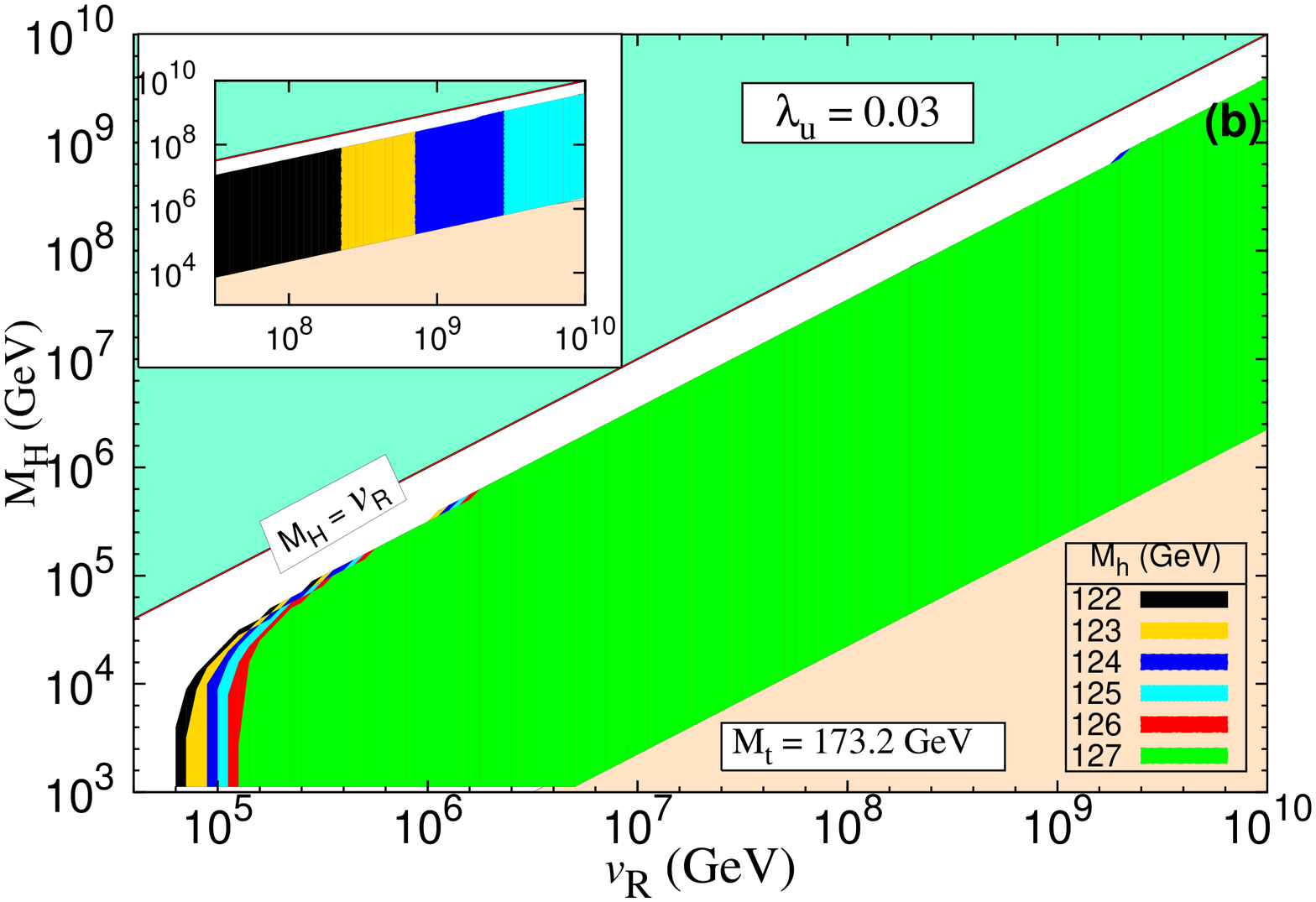} 
 \caption{Compatibility for stable vacuum in $v_R$ and heavy scalar $M_{H}$ allowed region in LR model 
 with triplet scalar. Each color represents a particular set of light Higgs mass ($M_h$) and top mass ($M_t$) in respective plot. 
 In figure (a) Higgs mass is fixed at 125 GeV for different top quark mass where as, in figure (b) top quark mass 
 is fixed at 173.2 GeV and Higgs mass is varying. Upper-left region (shaded with light blue) above the line $M_{H}=v_R$ is 
 disallowed since  quartic couplings are non-perturbative in this domain. 
 Lower-right region (shaded with light pink) 
 quartic coupling related with heavy scalar mass becomes extremely small ($\le \mathcal{O}(10^{-7})$). 
 We choose universal quartic coupling $\lambda_u$ fixed at 0.03. 
 Inset to both figures show the higher $v_R$ scale where color patches terminate, representing the very scale where in 
fact Standard Model breaks down for a particular Higgs mass or top quark mass at one loop.}
 \label{fig:LRT-vR-MH10}
 \end{center}
 \end{figure}
 
In figure~\ref{fig:LRT-vR-MH10} we check the compatibility for the stable vacuum in left-right 
symmetric breaking scale $v_R$ and heavy scalar $M_{H}$ allowed region in LR model 
 with triplet scalar. Each color represents a particular set of light Higgs mass ($M_h$) and top mass ($M_t$) in respective plot. 
 In figure~\ref{fig:LRT-vR-MH10}(a) Higgs mass is fixed at 125 GeV and top quark mass is varying from 170 GeV to 175 GeV where as, in figure~\ref{fig:LRT-vR-MH10}(b) top quark mass 
 is fixed at 173.2 GeV and Higgs mass is varying from 122 GeV to 127 GeV. Upper-left region (shaded with light blue) above the line $M_{H}=v_R$ is 
 disallowed since  quartic couplings are non-perturbative in this domain. 
 The blank (white) strip is also ruled out as the value of the couplings in this region is such that they become 
 non-perturbative before reaching the Planck scale. Lower-right region (shaded with light pink) 
 quartic coupling related with heavy Higgs mass becomes extremely small ($\le \mathcal{O}(10^{-7})$). 
 We choose universal quartic coupling $\la_2$, $\la_3$, $\la_4$, $\la_8$, $\la_9$, $\la_{10}$, $\la_{11} = \lambda_u$ fixed at 0.03.
 This choice of $\lambda_u$ allows only $v_R \geq$ 100 TeV which can be inferred from figure~\ref{fig:LRT-vR-l-uinversal}.
 Inset to both figures shows the higher $v_R$ scale where color patches terminate, representing the very scale where in 
fact Standard Model breaks down for a particular Higgs mass or top quark mass at one loop.


\subsubsection{LR Model with Doublet Scalars} \label{sec:vac_LR_doublet}

Using the similar technique used in previous section we depicted all the multiple field directions of the potential and the corresponding stability criteria in appendix~\ref{app:vac_LR_doublet}. We find the non-trivial vacuum stability conditions which read as
\be
\la_1 > 0 ,\hskip 15pt 2\beta_1+f_1 > 0, \hskip 15pt 2\beta_1-f_1>0.
\ee
\begin{figure}[tb]
\begin{center}
\includegraphics[width=7cm,angle=-90]{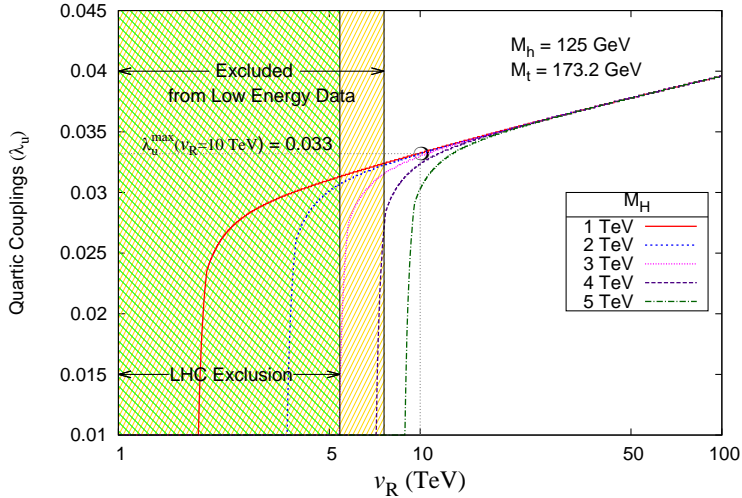}
 \caption{Constraints on universal quartic coupling $\la_u$ ($\equiv$ $\la_2$, -$\la_3$) for 
 LR model with doublet scalars in low $v_R$ region for different set of heavy scalar masses $M_H$. 
 Yellow Shaded region is disallowed from low energy data ($M_{W_R}>3.5$ TeV) and 
 green shaded region is excluded from direct search at LHC ($M_{W_{R}}>2.5$ TeV).}
 \label{fig:LRD-vR-l-universal}
 \end{center}
 \end{figure}
 We have also noted the required RGEs for our analysis in appendix~\ref{app:RG_LR_doublet} \cite{Rothstein:1990qx_LRT}.
 In figure~\ref{fig:LRD-vR-l-universal} we constrain universal quartic coupling $\la_u$ ($\equiv$ $\la_2$, -$\la_3$) 
 for LR model with doublet scalars in low $v_R$ region for different set of heavy scalar masses $M_H$. 
 Similar to the previous case, yellow shaded region in the plot is disallowed from low energy data ($M_{W_R}>3.5$ TeV) and 
 green shaded region is excluded from direct search at LHC ($M_{W_{R}}>2.5$ TeV).
 
 As we noticed at figure~\ref{fig:LRD-vR-l-universal}, for any particular heavy scalar mass ($M_H$), 
 universal quartic coupling $\la_u$ is disallowed above the corresponding line. For example, as seen from the plot, 
 maximum allowed value of the universal quartic coupling is  0.033
 if one consider LR breaking scale at 10 TeV and heavy scalar mass at 1 TeV. 
 As before, allowed maximum quartic coupling is lowered for heavier scalar.

\begin{figure}[tb]
\begin{center}
\includegraphics[width=5cm,angle=-90]{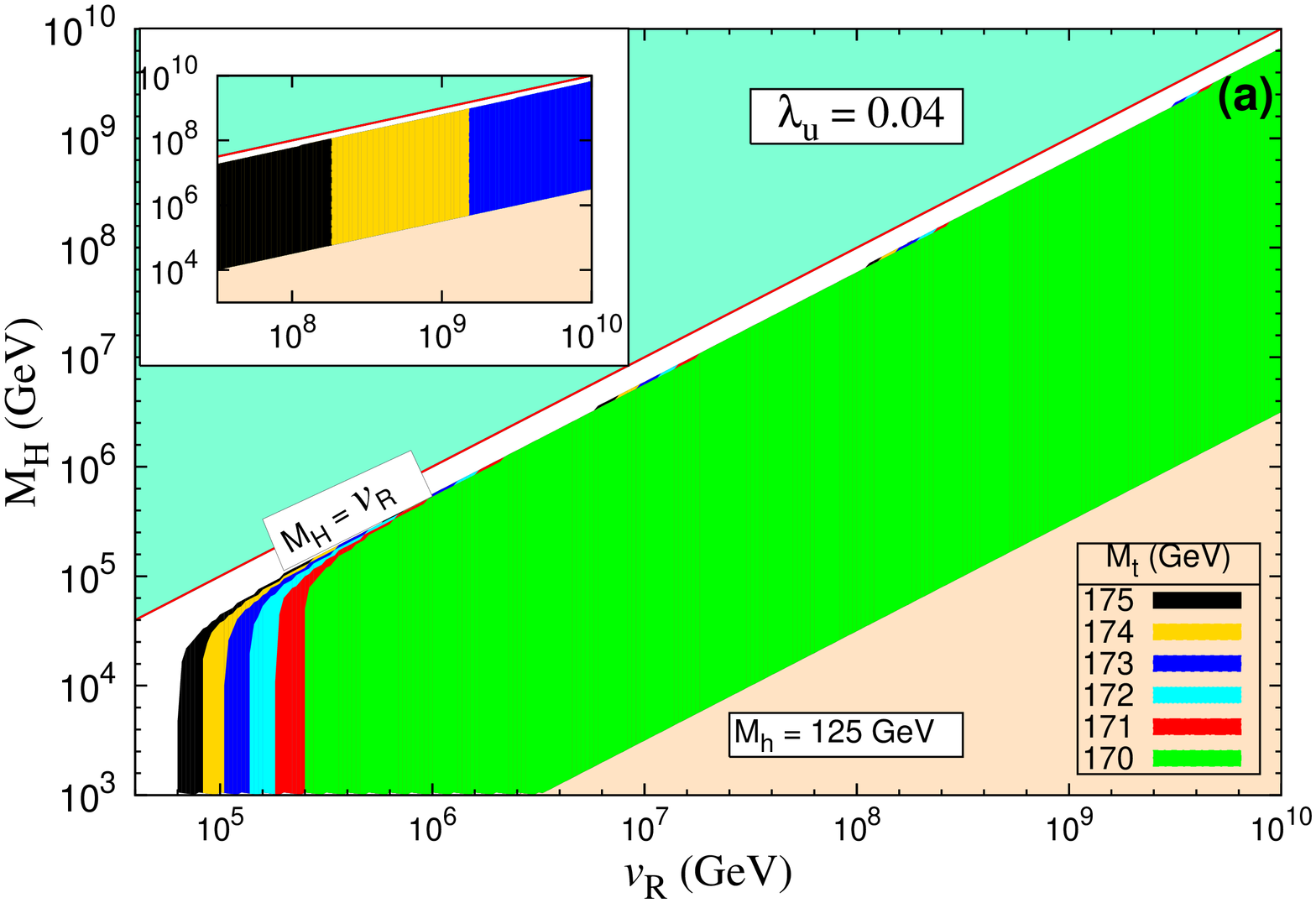}
\includegraphics[width=5cm,angle=-90]{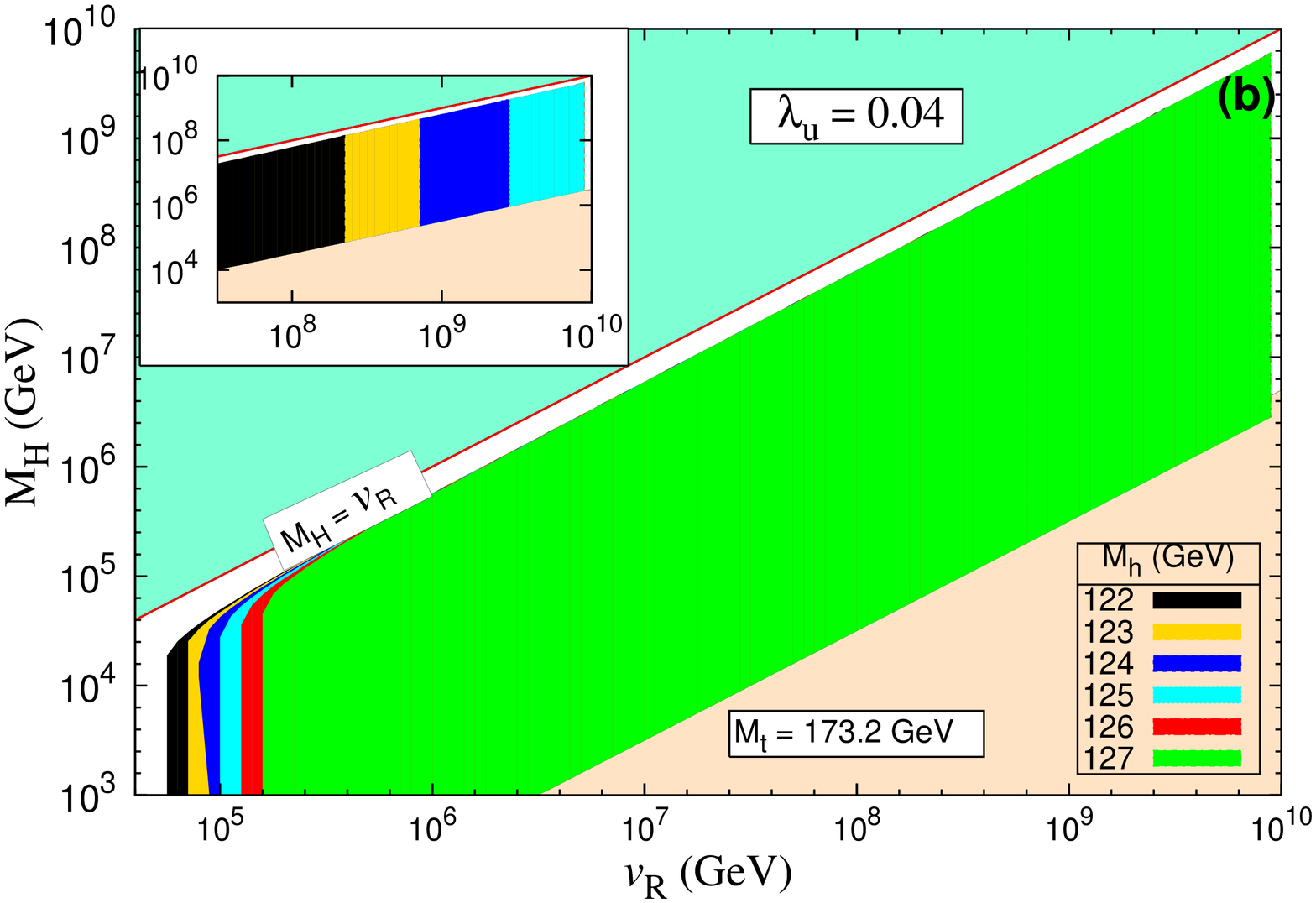} 
 \caption{Compatibility for stable vacuum in $v_R$ and heavy Higgs $M_{H}$ 
 allowed region in LR model with doublet scalars. Each color represents a particular set of light Higgs mass ($M_h$) 
 and top mass ($M_t$) in respective plot. In figure (a) Higgs mass is fixed at 125 GeV and top quark mass is varying, 
 where as, in figure (b) top quark mass is fixed at 173.2 GeV and Higgs mass is varying. 
 Upper-left region (shaded with light blue) above the line $M_{H}=v_R$ is disallowed since  quartic couplings are non-perturbative at the low scale  itself in this domain. 
 Lower-right region (shaded with light pink) quartic coupling related with heavy Higgs mass becomes extremely small ($\le \mathcal{O}(10^{-7})$). 
 We choose universal quartic coupling $\lambda_u$ fixed at 0.04. 
 Inset to both figures shows the higher $v_R$ scale where color patches terminate, 
 representing the very scale where in fact Standard Model breaks down for a 
particular Higgs mass or top quark mass at one loop.}
 \label{fig:LRD-vR-MH10}
 \end{center}
 \end{figure}

In figure~\ref{fig:LRD-vR-MH10} we check the compatibility for stable vacuum in  $v_R$ and heavy scalar $M_{H}$ 
 allowed region in LR model with doublet scalars. Each color represents a particular set of light Higgs mass ($M_h$) 
 and top mass ($M_t$) in respective plot. In figure~\ref{fig:LRD-vR-MH10}(a) Higgs mass is fixed at 125 GeV and top quark mass is varying 
 from 170 GeV to 175 GeV where as, in figure~\ref{fig:LRD-vR-MH10}(b) top quark mass is fixed at 173.2 GeV and Higgs mass is varying from 122 GeV to 127 GeV. 
 Upper-left region (shaded with light blue) above the line $M_{H}=v_R$ is disallowed since  quartic couplings are non-perturbative at the low scale 
 itself in this domain. The blank (white) strip is also ruled out as the value of the couplings in this region is such that they become 
 non-perturbative before reaching the Planck scale.  
 Lower-right region (shaded with light pink) quartic coupling related with heavy Higgs mass becomes extremely small ($\le \mathcal{O}(10^{-7})$). 
 We choose universal quartic coupling $\lambda_1 = - \lambda_2 = \lambda_u$ fixed at 0.04. 
 Here, the choice of $\lambda_u$ allows only $v_R \geq$ 100 TeV.
 Inset to both figures show the higher $v_R$ scale where color patches terminate, 
 representing the very scale where in fact Standard Model breaks down for a 
particular Higgs mass or top quark mass at one loop.

\section{Conclusions}
We have noted that one needs to study the scalar potential to understand the structure of the vacuum and its compatibility with 
successful spontaneous symmetry breaking. In addition, the perturbativity (triviality) of the couplings also plays a crucial role.
We have analysed the structure of the scalar potentials of $B-L$ extended models -- namely, SM$\otimes U(1)_{B-L}$ and left-right symmetry -- with different scalar representations. 
We have computed the criteria for the potential to be bounded from below, i.e., the conditions for vacuum stability. We also performed the renormalization group
evolutions of the parameters (couplings) of these models at the one loop level with proper matching conditions. We have shown how the phenomenologically 
unaccessible couplings can be constrained for different choices of scales of new physics. They in turn also affect the RGEs of the other couplings. 
We have noted that the new physics effects must be switched on before the SM vacuum face the instability. This helps the vacuum stability 
of the full scalar potential and achieve a consistent spontaneous symmetry breaking. We have analyzed these aspects by varying the Higgs and top quark mass
over their allowed ranges. In summary, it is meaningful to mention that more precise knowledge of the SM parameters, like Higgs mass, top quark mass and 
strong coupling will constrain the parameters (couplings, masses, scales) of new physics and that might direct us towards the correct theory for  beyond standard model physics. In principle one can study the left-right symmetric 
models including the radiative correction in the scalar potential and use the Coleman-Weinberg mechanism, e.g.,\cite{Holthausen:2009uc_LRD} has considered the scenario and calculated the flat directions using one-loop effective potential. This will certainly change the correlations among the parameters of the scalar potential leading to stable vacuum.
While submitting our paper there appeared \cite{u1b-l_recent} where the vacuum stability for SM$\otimes U(1)_{B-L}$ has been discussed. The view points of our analysis is quite different from this work.

\section*{Acknowledgements}
Work of JC is supported by Department of Science \& Technology, Government of INDIA under the Grant Agreement number IFA12-PH-34 (INSPIRE Faculty Award). Authors want to thank Srubabati Goswami and Namit Mahajan for useful discussions.

\newpage

\appendix

\section{Scalar Potential for Different Models} \label{app:Higgs_potential}

\subsection{$U(1)_{B-L}$ Model}\label{app:Higgs_potential_B-L}

\be \label{eq:BL-potential-1}
V(\Phi,S )=m^2\Phi^{\dagger}\Phi + \mu ^2\mid S \mid ^2 + \lambda_1 (\Phi^{\dagger}\Phi)^2 +\lambda_2 \mid S \mid ^4 + 
\lambda_3\, \Phi^{\dagger}\Phi\mid S \mid ^2.  
\ee

\subsection{LR model with triplet scalars}\label{app:Higgs_potential_LRT}

Most general form of the scalar potential can be written as in~\cite{Rothstein:1990qx_LRT} 

\beas\label{eq:LR_triplet_pot}
 && V_{LRT}(\Phi,\Delta_L,\Delta_R) = \\
    &-& \mu_1^2\bigg\{\Tr\big[\Phi^\dagger \Phi\big]\bigg\} 
    - \mu_2^2\bigg\{\Tr\big[\tilde{\Phi}\Phi^\dagger\big]+\Tr\big[\tilde{\Phi}^\dagger \Phi\big]\bigg\} 
    - \mu_3^2\bigg\{ \Tr\big[\Delta_L^\dagger \Delta_L\big]+\Tr\big[\Delta_R^\dagger \Delta_R\big] \bigg\}  \\
    &+& \lambda_1\bigg\{\Big(\Tr\big[\Phi^\dagger \Phi\big]\Big)^2\bigg\} +
    \lambda_2\bigg\{ \Big(\Tr\big[\tilde{\Phi}\Phi^\dagger\big]\Big)^2+\Big(\Tr\big[\tilde{\Phi}^\dagger \Phi\big]\Big)^2 \bigg\} 
    + \lambda_3\bigg\{\Tr\big[\tilde{\Phi}\Phi^\dagger\big]\Tr\big[\tilde{\Phi}^\dagger \Phi\big] \bigg\}\\
    &+& \lambda_4 \bigg\{ \Tr\big[\Phi^\dagger \Phi\big]\Big(\Tr\big[\tilde{\Phi}\Phi^\dagger\big]
    +\Tr\big[\tilde{\Phi}^\dagger \Phi\big]\Big) \bigg\}
    +\lambda_5\bigg\{ \Big(\Tr\big[\Delta_L \Delta_L^\dagger\big]\Big)^2+\Big(\Delta_R \Delta_R^\dagger\Big)^2 \bigg\} \\
    &+& \lambda_6 \bigg\{\Tr\big[\Delta_L \Delta_L\big]\;\Tr\big[\Delta_L^\dagger \Delta_L^\dagger\big]
    +\Tr\big[\Delta_R \Delta_R\big]\;\Tr\big[\Delta_R^\dagger \Delta_R^\dagger\big]  \bigg\}
    + \lambda_7 \bigg\{\Tr\big[\Delta_L\Delta_L^\dagger\big]\;\Tr\big[\Delta_R\Delta_R^\dagger\big]\bigg\} \\
    &+& \lambda_8[\Delta_L\Delta_L^\dagger\big] \bigg\{\Tr\big[\Delta_L\Delta_L^\dagger\big]\;
    \Tr\big[\Delta_R\Delta_R^\dagger\big] \bigg\}
    + \lambda_9 \bigg\{\Tr\big[\Phi^\dagger \Phi\big]\Big(\Tr\big[\Delta_L\Delta_L^\dagger\big]
    +\Tr\big[\Delta_R\Delta_R^\dagger\big]\Big)\bigg\} \\
    &+& (\lambda_{10}+i\,\lambda_{11}) 
    \bigg\{\Tr\big[\Phi\tilde{\Phi}^\dagger\big]\Tr\big[\Delta_R\Delta_R^\dagger\big] 
    + \Tr\big[\Phi^\dagger\tilde{\Phi}\big]\Tr\big[\Delta_L\Delta_L^\dagger\big]\bigg\} \\
    &+& (\lambda_{10}-i\,\lambda_{11}) 
    \bigg\{\Tr\big[\Phi^\dagger\tilde{\Phi}\big]\Tr\big[\Delta_R\Delta_R^\dagger\big] 
    + \Tr\big[\tilde{\Phi}^\dagger\Phi\big]\Tr\big[\Delta_L\Delta_L^\dagger\big]\bigg\} \\
    &+& \lambda_{12} \bigg\{ \Tr\big[\Phi \Phi^\dagger \Delta_L \Delta_L^\dagger\big]
    +\Tr\big[\Phi^\dagger \Phi \Delta_R \Delta_R^\dagger\big]\bigg\}
    + \lambda_{13} \bigg\{\Tr\big[\Phi\Delta_R\Phi^\dagger\Delta_L^\dagger\big]
    +\Tr\big[\Phi^\dagger\Delta_L\Phi\Delta_R^\dagger\big] \bigg\} \\
    &+& \lambda_{14}\bigg\{\Tr\big[\tilde{\Phi}\Delta_R\Phi^\dagger\Delta_L^\dagger\big]
    +\Tr\big[\tilde{\Phi}^\dagger\Delta_L\Phi\Delta_R^\dagger\big]\bigg\} 
    + \lambda_{15} \bigg\{\Tr\big[\Phi\Delta_R\tilde{\Phi}^\dagger\Delta_L^\dagger\big]
    +\Tr\big[\Phi^\dagger\Delta_L\tilde{\Phi}\Delta_R^\dagger\big]\bigg\},
\eeas
where all the coupling constants are real. 

\subsection{LR model with doublet scalars}\label{app:Higgs_potential_LRD}

Scalar potential for LR model with doublet scalars can be written as:
\beas\label{eq:LR_doublet_pot}
V_{LRD} (\Phi,H_L,H_R) &=&  4 \la_1 \Big(\Tr[\Phi^\dagger\Phi]\Big)^2 +
4 \la_2 \Big(\Tr[\Phi^\dagger \tilde{\Phi}]+\Tr[\Phi\tilde{\Phi}^\dagger]\Big)^2 +
4 \la_3 \Big(\Tr[\Phi^\dagger \tilde{\Phi}]-\Tr[\Phi\tilde{\Phi}^\dagger]\Big)^2 \\
&& + \frac{\kappa_1}{2} \Big(H_L^\dagger H_L+H_R^\dagger H_R\Big)^2 +
\frac{\kappa_2}{2} \Big(H_L^\dagger H_L-H_R^\dagger H_R\Big)^2 \\
&& + \beta_1 \Big(\Tr[\Phi^\dagger \tilde{\Phi}]+\Tr[\Phi\tilde{\Phi}^\dagger]\Big)\Big(H_L^\dagger H_L+H_R^\dagger H_R\Big)\\
&& + f_1\Big(H_L^\dagger\big(\tilde{\Phi}\tilde{\Phi}^\dagger-\Phi \Phi^\dagger) H_L
-H_R^\dagger \big(\Phi^\dagger\Phi-\tilde{\Phi}^\dagger \tilde{\Phi} \big)H_R \Big).
\eeas


\section{Calculation of non-trivial Vacuum Stability Conditions}
\label{app:Higgs_potential_VS}

Here we have gathered the structure of the scalar potential in 2,3 and 4-field directions. We have calculated vacuum stability conditions from these fields direction 
keeping in mind that the conditions should cover most of the parameter space spanned by the quartic couplings.

\subsection{$U(1)_{B-L}$ Model} 
\label{app:vac_B-L}
For $U(1)_{B-L}$ model the potential has a simple structure has a simple structure and the stability cositions can be calculated easily.
The quartic potential has the form $$ \la_1\,|\Phi|^4 + \la_2\,|S|^4+\la_3|\Phi|^2\,|S|^2,$$ and we can easily write this potential as 
$$\left(\sqrt{\la_1}\,|\Phi|^2+\frac{\la_3}{2\sqrt{\la_1}}\,|S|^2\right)^2 + \left(\la_2-\frac{\la_3^2}{4\la_1}\right)|S|^4. $$
Clearly the above equation is positive definite if 
\beas
\la_1>0\;,\;\;\la_2>0,\\ 
4\la_1\la_2-\la_3^2>0.
\eeas
These are the non-trivial vacuum stability conditions with $\lambda_3<0$. The trivial boundary conditions are when all the $\lambda_{1,2,3}>0$.


\subsection{LR Model with Triplet Scalars }
\label{app:vac_LR_triplet}
Absence of any tachyonic pseudoscalar modes  imposes the condition 
$\la_{12} > 0$.

\subsubsection{2 Field Directions and Stability Conditions}
\beas
^{2F}V_1(\phi_1^0\,,\,\phi_1^+)    &=& \la_1 \big({\phi_1^0}^2 + {\phi_1^+}^2 \big)^2 \\
^{2F}V_2(\phi_1^0\,,\,\delta^{0})  &=& \la_5 \; {\delta^{0}}^4 + \la_1\;{\phi_1^0}^4 \\
^{2F}V_3(\phi_1^0\,,\,\delta^{+})  &=& (\la_5 + \la_6) \; {\delta^+}^4 + \la_1\;{\phi_1^0}^4 + \frac{1}{2}(\la_{12}+2\la_9)\;{\delta^{+}}^2 {\phi_1^0}^2 \\
^{2F}V_4(\phi_1^0\,,\,\delta^{++}) &=& \la_5\;{\delta^{++}}^4 + \la_1\;{\phi_1^0}^4 + \la_{12}\;{\delta^{++}}^2 {\phi_1^0}^2\\
^{2F}V_5(\phi_1^+\,,\,\delta^{0})  &=& \la_5 \; {\delta^0}^4 + \la_1\;{\phi_1^+}^4 +\la_{12}\;{\delta^{0}}^2 {\phi_1^+}^2\\
^{2F}V_6(\phi_1^+\,,\,\delta^{+})  &=& (\la_5 + \la_6) \; {\delta^+}^4 + \la_1\;{\phi_1^+}^4 + \frac{1}{2}(\la_{12}+2\la_9)\;{\delta^{+}}^2 {\phi_1^+}^2 \\
^{2F}V_7(\phi_1^+\,,\,\delta^{++}) &=& \la_5 \; {\delta^{++}}^4 + \la_1\;{\phi_1^0}^4 \\
^{2F}V_8(\delta^0\,,\,\delta^{+})  &=& \la_5 \big({\delta^0}^2 + {\delta^{+}}^2 \big)^2 + \la_6\; {\delta^{+}}^4 \\
^{2F}V_9(\delta^0\,,\,\delta^{++}) &=& \la_5 \big({\delta^0}^2 + {\delta^{++}}^2 \big)^2 + 4 \la_6\; {\delta^{+}}^2{\delta^{0}}^2 \\
^{2F}V_{10}(\delta^+\,,\,\delta^{++}) &=& \la_5 \big({\delta^+}^2 + {\delta^{++}}^2 \big)^2 + \la_6\; {\delta^{+}}^4 \\
\eeas

{\bf Stability conditions}
\beas
^{2F}V_1 & \longrightarrow & \la_1 > 0 \\
^{2F}V_2,\;^{2F}V_4,\;^{2F}V_5,\;^{2F}V_7 & \longrightarrow & \la_1 > 0\,;\;\; \la_5>0 \\
^{2F}V_3,\;^{2F}V_6& \longrightarrow & \la_1 > 0\,;\;\; \la_5+\la_6>0 \\
^{2F}V_8,\;^{2F}V_9,\;^{2F}V_{10} & \longrightarrow & \la_5 > 0\,;\;\; \la_5+\la_6>0 
\eeas

\subsubsection{3 Field Directions and Stability Conditions}
\beas
^{3F}V_1 (\phi_1^0\,,\,\phi_1^+\,,\,\delta^{0})    &=&  \la_1\; \big({\phi_1^0}^2+{\phi_1^+}^2 \big)^2 +\la_5\;{\delta^0}^2 +\la_{12}\;{\delta^0}^2{\phi_1^0}^2\\
^{3F}V_2  (\phi_1^0\,,\,\phi_1^{+}\,,\,\delta^{+}) &=& \la_1\; \big({\phi_1^0}^2+{\phi_1^+}^2 \big)^2 + (\la_5+\la_6){\delta^+}^4 
                                                    +\frac{1}{2}(\la_{12}+2\la_9)\;\big({\phi_1^0}^2+{\phi_1^+}^2 \big) {\delta^+}^2\\
^{3F}V_3 (\phi_1^0\,,\,\phi_1^{+}\,,\,\delta^{++}) &=&  \la_1\; \big({\phi_1^0}^2+{\phi_1^+}^2 \big)^2 + \la_5\;{\delta^{++}}^4 + \la_{12}\;{\phi_1^0}^2 {\delta^{++}}^2\\
^{3F}V_4 (\phi_1^0\,,\,\delta^{0}\,,\,\delta^{+})  &=&  \la_1\;{\phi_1^0}^4+\la_5\; \big({\delta^0}^2+{\delta^{+}}^2\big)^2 + \la_6 {\delta^+}^4 +
                                                        \frac{1}{2}(\la_{12}+2\la_9)\;{\phi_1^0}^2 {\delta^+}^2  \\
^{3F}V_5 (\phi_1^0\,,\,\delta^{0}\,,\,\delta^{++}) &=&  \la_5\; \big({\delta^0}^2+{\delta^{++}}^2\big)^2 + \la_1 \;{\phi_1^0}^4 
                                                        + 4 \la_6\; {\delta_0}^2{\delta^{++}}^2 + \la_{12}\; {\delta^{++}}^2\, {\phi_1^0}^2 + 2\,\la_9\,\delta^0\,\delta^{++} \, {\phi_1^0}^2   \\
^{3F}V_6 (\phi_1^0\,,\,\delta^{+}\,,\,\delta^{++}) &=&  \la_1\;{\phi_1^0}^4+\la_5\; \big({\delta^{++}}^2+{\delta^{+}}^2\big)^2 + \la_6\; {\delta^{+}}^4 +
                                                        \frac{1}{2}\la_{12}\;{\phi_1^0}^2 (2{\delta^{++}}^2+{\delta^{+}}^2) + \la_9\, {\delta^+}^2 {\phi_1^0}^2\\
^{3F}V_7 (\phi_1^+\,,\,\delta^{0}\,,\,\delta^{+})  &=&  \la_1\;{\phi_1^+}^4+\la_5\; \big({\delta^0}^2+{\delta^{+}}^2\big)^2 + \la_6\; {\delta^{+}}^4 +
                                                        \frac{1}{2}\la_{12}\;{\phi_1^+}^2 (2{\delta^0}^2+{\delta^{+}}^2) + \la_9\, {\delta^+}^2 {\phi_1^+}^2 \\
^{3F}V_8 (\phi_1^+\,,\,\delta^{0}\,,\,\delta^{++}) &=&  \la_5\; \big({\delta^0}^2+{\delta^{++}}^2\big)^2 + \la_1 \;{\phi_1^+}^4 
                                                        + 4 \la_6\; {\delta_0}^2{\delta^{++}}^2 + \la_{12}\; {\delta^0}^2\, {\phi_1^+}^2 + 2\,\la_9\,\delta^0\,\delta^{++} \, {\phi_1^+}^2   \\
^{3F}V_9 (\phi_1^+\,,\,\delta^{+}\,,\,\delta^{++}) &=&  \la_1\;{\phi_1^+}^4+\la_5\; \big({\delta^+}^2+{\delta^{++}}^2\big)^2 + \la_6 {\delta^{+}}^4 +
                                                        \frac{1}{2}(\la_{12}+2\,\la_9)\;{\phi_1^+}^2 {\delta^{+}}^2   \\
^{3F}V_{10} (\delta^0\,,\,\delta^{+}\,,\,\delta^{++}) &=&  \la_5\; \big({\delta^0}^2+{\delta^{+}}^2+{\delta^{++}}^2\big)^2 
                                                        +\la_6\; \big({\delta^{+}}^2+2\delta^0\delta^{++}\big)^2 \\
\eeas

{\bf Stability conditions}
\beas
^{3F}V_1,\;^{3F}V_3 & \longrightarrow &  \la_1 > 0\,;\;\; \la_5>0\\
^{3F}V_2 & \longrightarrow & \la_1 > 0\,;\;\; \la_5+\la_6>0 \\
^{3F}V_4,\;^{3F}V_6,\;^{3F}V_7,\;^{3F}V_9 & \longrightarrow & \la_1 > 0\,;\;\; \la_5>0\,;\;\; \la_5+\la_6>0\\
^{3F}V_5,\;^{3F}V_8  & \longrightarrow & \la_1 > 0\,;\;\; \la_5>0\,;\;\; \la_5+2\la_6>0\\
^{3F}V_{10} & \longrightarrow & \la_5>0\,;\;\; \la_5+\la_6>0\,;\;\; \la_5+2\la_6>0
\eeas

\subsubsection{4 Field Directions and Stability Conditions}
\beas
^{4F}V_{1} (\phi_1^0\,,\,\phi_1^+\,,\,\delta^{0}\,,\,\delta^{+}) &=&  \la_5\; \big({\delta^0}^2+{\delta^{+}}^2\big)^2 +\la_6\;{\delta^{+}}^4
                                                                        +\la_1\; \big({\phi_1^0}^2+{\phi_1^+}^2 \big)^2 \\
                                                                        && + \frac{1}{2}\la_{12}\; \big(2{\delta^{0}}^2{\phi_1^+}^2 
                                                                        + 2\sqrt{2}\phi_1^0\phi_1^+\delta^0\delta^{+}+{\delta^+}^2\big({\phi_1^0}^2+{\phi_1^+}^2 \big)  \big)\\&& + \la_9 {\delta^+}^2\left({\phi_1^0}^2+{\phi_1^+}^2\right) \\
^{4F}V_{2} (\phi_1^0\,,\,\phi_1^+\,,\,\delta^{0}\,,\,\delta^{++}) &=&  \la_5\; \big({\delta^0}^2+{\delta^{++}}^2\big)^2 + 4\la_6\;{\delta^0}^2{\delta^{++}}^2
                                                                        +\la_1\; \big({\phi_1^0}^2+{\phi_1^+}^2 \big)^2  \\
                                                                        && + \la_{12} \big( {\delta^{++}}^2{\phi_1^0}^2+{\delta^0}^2{\phi_1^+}^2 \big) +                                              2\la_9\,\delta^0\,\delta^{++}\,\left({\phi_1^0}^2+{\phi_1^+}^2\right)  \\
^{4F}V_{3} (\phi_1^0\,,\,\phi_1^+\,,\,\delta^{+}\,,\,\delta^{++}) &=&  \la_5\; \big({\delta^+}^2+{\delta^{++}}^2\big)^2 +\la_6\;{\delta^{+}}^4
                                                                        +\la_1\; \big({\phi_1^0}^2+{\phi_1^+}^2 \big)^2 \\
                                                                        && + \frac{1}{2}\la_{12}\; \big(2{\delta^{++}}^2{\phi_1^0}^2-2\sqrt{2}\phi_1^0\phi_1^
                                                                        +\delta^+\delta^{++}+{\delta^+}^2\big({\phi_1^0}^2+{\phi_1^+}^2 \big)  \big) \\&& + \la_9 {\delta^+}^2\left({\phi_1^0}^2+{\phi_1^+}^2\right)\\
^{4F}V_{4} (\phi_1^0\,,\,\delta^0\,,\,\delta^{+}\,,\,\delta^{++}) &=&  \la_5\; \big({\delta^0}^2+{\delta^{+}}^2+{\delta^{++}}^2\big)^2 
                                                                        +\la_6\; \big({\delta^{+}}^2+2\delta^0\delta^{++}\big)^2+\la_1\;{\phi_1^0}^4 \\
                                                                        && +\frac{1}{2}\la_{12}\;{\phi_1^0}^2 (2{\delta^0}^2+{\delta^{+}}^2) + \la_9 {\phi_1^0}^2\left({\delta^+}^2+2 \delta^0\,\delta^{++}\right) \\
^{4F}V_{5} (\phi_1^+\,,\,\delta^0\,,\,\delta^{+}\,,\,\delta^{++}) &=&  \la_5\; \big({\delta^0}^2+{\delta^{+}}^2+{\delta^{++}}^2\big)^2 
                                                                        +\la_6\; \big({\delta^{+}}^2+2\delta^0\delta^{++}\big)^2+\la_1\;{\phi_1^+}^4 \\
                                                                        && +\frac{1}{2}\la_{12}\;{\phi_1^+}^2 (2{\delta^0}^2+{\delta^{+}}^2) + \la_9 {\phi_1^+}^2\left({\delta^+}^2+2 \delta^0\,\delta^{++}\right)
\eeas

{\bf Stability conditions}
\beas
^{4F}V_1 & \longrightarrow & \la_1 > 0\,;\;\;\la_5>0\,;\;\; \la_5+2\la_6>0 \\
^{4F}V_2 & \longrightarrow & \la_1 > 0\,;\;\;\la_5>0\,;\;\; \la_5+\la_6>0 \\
^{4F}V_3 & \longrightarrow & \la_1 > 0\,;\;\;\la_5>0\,;\;\; \la_5+\la_6>0\,;\;\; \la_{12}-2\sqrt{2\,\la_1\,\la_5} < 0 \\
^{4F}V_4,\;^{4F}V_5 & \longrightarrow & \la_1 > 0\,;\;\;\la_5>0\,;\;\; \la_5+\la_6>0\,;\;\; \la_5+2\la_6>0 
\eeas


\subsection{LR Model with Doublet Scalars}
\label{app:vac_LR_doublet}
\subsubsection{2 Field Directions and Stability Conditions}
\beas
 ^{2F}V_1(\phi_1^0\,,\,\phi_1^+)    &=& \la_1\;\left({\phi_1^0}^2+{\phi_1^+}^2\right)^2 \\
 ^{2F}V_2(\phi_1^+\,,\,h_R^+)       &=& \la_1\;{\phi_1^+}^4 + \frac{2\beta_1+f_1}{2}{h_R^+}^2 {\phi_1^+}^2 \\
 ^{2F}V_3(\phi_1^0\,,\,h_R^+)       &=& \la_1\;{\phi_1^0}^4 + \frac{2\beta_1-f_1}{2}{h_R^+}^2 {\phi_1^0}^2 \\ 
 ^{2F}V_4(\phi_1^+\,,\,h_R^0)       &=& \la_1\;{\phi_1^+}^4 + \frac{2\beta_1-f_1}{2}{h_R^0}^2 {\phi_1^+}^2 \\  
 ^{2F}V_5(\phi_1^0\,,\,h_R^0)       &=& \la_1\;{\phi_1^0}^4 + \frac{2\beta_1+f_1}{2}{h_R^0}^2 {\phi_1^0}^2 \\
\eeas

{\bf Stability conditions}

\beas
 ^{2F}V_1  & \longrightarrow & \la_1 > 0 \\
 ^{2F}V_2,\;^{2F}V_5 & \longrightarrow &  \la_1 > 0\,;\;\; 2\beta_1+f_1>0 \\
 ^{2F}V_3,\;^{2F}V_4 & \longrightarrow & \la_1 > 0\,;\;\; 2\beta_1-f_1>0 
\eeas

\subsubsection{3 Field Directions and Stability Conditions}
\beas
^{3F}V_1 (\phi_1^0\,,\,\phi_1^+\,,\,h_R^{0})  &=& \la_1 \left({\phi_1^0}^2+{\phi_1^+}^2\right)^2+
                                                  {h_R^0}^2\left(\beta_1({\phi_1^0}^2+{\phi_1^+}^2)+\frac{1}{2}f_1({\phi_1^0}^2-{\phi_1^+}^2)\right)\\
^{3F}V_2 (\phi_1^0\,,\,\phi_1^+\,,\,h_R^{+})  &=& \la_1 \left({\phi_1^0}^2+{\phi_1^+}^2\right)^2+
                                                  {h_R^+}^2\left(\beta_1({\phi_1^0}^2+{\phi_1^+}^2)+\frac{1}{2}f_1({\phi_1^+}^2-{\phi_1^0}^2)\right)\\
^{3F}V_3 (\phi_1^0\,,\,h_R^{0}\,,\,h_R^{+})  &=& \frac{1}{2}{\phi_1^0}^2\bigg(f_1\left({h_R^0}^2-{h_R^+}^2\right)
                                                  +2\beta_1\left({h_R^0}^2+{h_R^+}^2\right)+2\la_1{\phi_1^0}^2\bigg)\\
^{3F}V_3 (\phi_1^+\,,\,h_R^{0}\,,\,h_R^{+})  &=& \frac{1}{2}{\phi_1^+}^2\bigg(f_1\left({h_R^+}^2-{h_R^0}^2\right)
                                                  +2\beta_1\left({h_R^0}^2+{h_R^+}^2\right)+2\la_1{\phi_1^+}^2\bigg)\\
\eeas

{\bf Stability conditions}
\be
 ^{3F}V_1,\;^{3F}V_2,\;^{3F}V_3,\;^{3F}V_4  \longrightarrow \la_1 > 0\,;\;\; 2\beta_1+f_1>0\,;\;\; 2\beta_1-f_1>0 \nonumber
\ee

\subsubsection{4 Field Directions and Stability Conditions}

\beas
^{4F}V_1 (\phi_1^0\,,\,\phi_1^+\,,\,h_R^{0}\,,\,h_R^{+})  = \frac{1}{2}\Bigg(&& f_1\left(h_R^+(\phi_1^0-\phi_1^+)+h_R^0(\phi_1^0+\phi_1^+)\right)
                                                               \left(h_R^0(\phi_1^0-\phi_1^+)-h_R^+(\phi_1^0+\phi_1^+)\right) \\
                                                          && +2({\phi_1^0}^2+{\phi_1^+}^2) \left(({h_R^0}^2+{h_R^+}^2)\beta_1+\la_1({\phi_1^0}^2+{\phi_1^+}^2)\right)\Bigg)
\eeas

{\bf Stability conditions}
\be
 ^{4F}V_1  \longrightarrow \la_1 > 0\,;\;\; 2\beta_1+f_1>0\,;\;\; 2\beta_1-f_1>0 \nonumber
\ee


\section{Renormalization Group Evolution Equations}
\subsection{Standard Model RGEs}
For Standard Model we have used renormalization group evolution equations from \cite{Holthausen:2011aa_smRGE} with matching conditions for top Yukawa and 
Higgs quartic coupling at their pole masses.

\subsection{$U(1)_{B-L}$ Model} \label{app:RG_BL}

\subsubsection*{Gauge RG Equations}

Renormalization group equations for $SU(3)_C$ and $SU(2)_L$ gauge couplings $g_3$ and $g_2$:
\beas
{16\pi^2}\frac{d}{dt}g_3 &=& g_3^3\bigg[-1+\frac{4}{3}n_g\bigg] =  \frac{g_3^3}{16\pi^2} \bigg[-7\bigg]\\
{16\pi^2}\frac{d}{dt}g_2 &=& g_2^3\bigg[-\frac{22}{3}+\frac{4}{3}n_g+\frac{1}{6}\bigg] =  \frac{g_2^3}{16\pi^2} \bigg[-\frac{19}{6}\bigg]
\eeas
where $n_g$ is number of generations.

\noindent Renormalization group equations for Abelian gauge couplings $g_1,\;g_{_{B-L}} \;$and$\;\;\widetilde{g}$:
\beas
{16\pi^2}\frac{d}{dt}g_1  &=&  \bigg[ \frac{41}{6}\; g^3_1 \bigg]\\
{16\pi^2}\frac{d}{dt}g_{_{B-L}} &=&  \bigg[ 12\; g_{_{B-L}}^3+ \frac{32}{3} g_{_{B-L}}\widetilde{g}+\frac{41}{6} g_{_{B-L}}\widetilde{g}^2 \bigg]\\
{16\pi^2}\frac{d}{dt}\widetilde{g}  &=& \bigg[ \frac{41}{6} \;\widetilde{g}(\widetilde{g}^2+2 g^2_1)+
 \frac{32}{3}\; g_{_{B-L}} (\widetilde{g}^2+ g^2_1)+12\; g_{_{B-L}}^2\widetilde{g}\bigg]
\eeas

\subsubsection*{Fermion RG Equations}
RG evolution equation for top quark Yukawa coupling $Y_t$:
\bes
{16\pi^2}\frac{d}{dt}Y_t  = {Y_t}\bigg[\frac{9}{2}Y^2_t - 8 g^2_3 -\frac{9}{4}g^2_2-
\frac{17}{12}g^2_1-\frac{17}{12}\widetilde{g}^2-\frac{2}{3}g_{_{B-L}}^2-\frac{5}{3}\widetilde{g}g_{_{B-L}}  \bigg]
\ees

\noindent In case of RH neutrinos RGEs we are considering degenerate RH neutrino Yukawa coupling and 
 we are in a basis where these couplings are diagonal, then we have : 
\bes
{16\pi^2}\frac{d}{dt}y^h_i={y^h_i} \bigg[ 4(y^h_i)^2+2\;\Tr\big[(y^h)^2\big]- 6 g_{_{B-L}}^2 \bigg]
\ees

\subsubsection*{Scalar RG Equations}
RGEs for the scalar couplings $\la_1,\;\la_2$ and $\la_3$ are :

\beas
{16\pi^2}\frac{d}{dt} \la_1 &=&\bigg[ 24 \la_1^2 + \la_3^2 - 6Y_t^4 +\frac{9}{8} g^4_2 + \frac{3}{8}g^4_1 +
\frac{3}{4} g^2_2g^2_1 + \frac{3}{4} g^2_2 \widetilde{g}^2 + \frac{3}{4} g^2_1 \widetilde{g}^2 \nonumber  \\ 
&&+ \frac{3}{8} \widetilde{g}^4 + 12 \la_1 Y_t^2 - 9\la_1 g^2_2 - 3 \la_1 g^2_1 -3 \la_1 \widetilde{g}^2 \bigg] \\
{8\pi^2}\frac{d}{dt} \la_2 &=&\bigg[ 10\la_2^2 + \la_3^2 - \frac{1}{2}\Tr\big[(y^h)^4\big] +
48 g_{_{B-L}}^4+ 4\la_2\Tr\big[(y^h)^2\big] -24 \la_2 g_{_{B-L}}^2 \bigg]\\
{8\pi^2}\frac{d}{dt} \la_3 &=& {\la_3}\bigg[ 6\la_1+4\la_2+2\la_3+3 Y_t^2 -\frac{3}{4}(3 g^2_2 -g^2_1- \widetilde{g}^2)
+2\;\Tr\big[(y^h)^2\big]-12 g_{_{B-L}}^2 \bigg] \nonumber \\ && + 6\widetilde{g}^2g_{_{B-L}}^2\nonumber \\  
\eeas


\subsection{LR Model with Triplet Scalars } \label{app:RG_LR_triplet}

\subsubsection*{Gauge RG Equations}
\beas
{16\pi^2}\frac{d}{dt}g_3 &=&  {g_3^3} \bigg(-7\bigg)\\
{16\pi^2}\frac{d}{dt}g_{2} &=&  {g_{2}^3} \bigg(-\frac{15}{6}\bigg)\\
{16\pi^2}\frac{d}{dt}g_{_{B-L}}  &=& {g_{_{B-L}}^3} \bigg( \frac{28}{9} \bigg)
\eeas
Note that in our case $g_{_{2L}} = g_{_{2R}} = g_{_{2}}$. 
\subsubsection*{Fermion RG Equations}
\beas
{16\pi^2}\frac{d}{dt}Y_t &=&  \bigg[8 Y_t^3 - Y_t\big( \frac{2}{3} g^2_1-\frac{9}{2}g^2_2 - 8 g^2_3\big)\bigg]\\
{16\pi^2}\frac{d}{dt}Y^M_i &=&  \bigg[2Y^M_i\Big(-\frac{3}{4}g^2_1-\frac{9}{4}g^2_2\Big)+2Y^M_i\Tr\big[(Y^M)^2\big]+
6(Y^M_i)^3\bigg]
\eeas

\subsubsection*{Scalar RG Equations}
To write down scalar RG equations, We classified 15 scalar couplings into three categories depending on how they coupled with scalar fields.
\begin{itemize}
 \item {\bf Coefficients with $\Phi^4$}
 \beas
{16\pi^2}\frac{d}{dt}\la_1 &=& 32\la^2_1+\frac{5}{3}\la^2_{12}+\frac{1}{2}\la^2_{13}+2\la^2_{14}+64\la^2_2+16\la_1\la_3+16\la^2_3 \nonumber\\
&&+\;48\la_4^2+6\la_{12}\la_9+6\la_9^2+12\la_1 Y_t^2-6Y_t^4-18\la_1g_2^2+3g_2^4\\
{16\pi^2}\frac{d}{dt}\la_2 &=& 6(\la_{10}^2-\la{11}^2)+\frac{3}{2}\la_{14}\la_{15}+24\la_1\la_2+48\la_2\la_3\nonumber\\
&& +12\la_4^2+12\la_2Y_t^2-18\la_2 g_2^2\\
{16\pi^2}\frac{d}{dt}\la_3 &=&  12(\la_{10}^2+\la_{11}^2)-(\la_{12}^2-\la_{13}^2)-\frac{1}{2}\;(\la_{14}^2+\la_{15}^2)+128\la_2^2\nonumber\\
&& +24\la_1\la_3+16\la_3^2+24\la_4^2+12\la_3Y_t^2+3Y_t^4-18\la_3g_2^2+\frac{3}{2}g_2^2\\
{16\pi^2}\frac{d}{dt}\la_4 &=& 48\la_4(\la_1+2\la_2+\la_3)+6\la_{10}(2\la_9+\la_{12})\nonumber\\ 
&&+\frac{3}{2}\la_{13}(\la_{14}+\la_{15})+12\la_4Y_t^2-18\la_4g_2^2\\
\eeas

\item {\bf Coefficients with $\Delta^4$}
\beas
{16\pi^2}\frac{d}{dt}\la_5 &=& 28\la_5^2+16\la_6(\la_5+\la_6)+16(\la_{10}^2+\la_{11}^2)+2\la_{12}^2+3\la_7^2\nonumber\\
&& +4\la_9(\la_9+\la_{12})+2\la_5Y_t^2-16Y_t^4-12\la_5g_{_{B-L}}^2\nonumber\\
&& +6g_{_{B-L}}^4+12g_{_{B-L}}^2g_2^2-24\la_5g_2^2+9g_2^4\\
{16\pi^2}\frac{d}{dt}\la_6 &=& 12\la_6(\la_6+2\la_5-g_{_{B-L}}^2-2g_2^2)+12\la_8^2\nonumber\\
&&-\la_{12}^2+8Y_t^4+8\la_6Y_t^2-12g_{_{B-L}}^2g_2^2+3g_2^4\\
{16\pi^2}\frac{d}{dt}\la_7 &=& 4\la_7^2+16\la_7(2\la_5+\la_6)+32(\la_{10}^2-\la_{11}^2)+2(\la_{12}^2+\la_{13}^2)\nonumber\\
&& +4(\la_{14}^2+\la_{15}^2)+32\la_8^2+8\la_{12}\la_9+\la_9^2\nonumber\\
&& +8\la_7Y_t^2-12\la_7(g_{_{B-L}}^2+g_2^2)+12g_{_{B-L}}^4\\
{16\pi^2}\frac{d}{dt}\la_8 &=& \la_{13}^2+4\la_{14}\la_{15}+8\la_8(\la_5+5\la_6+\la_7+Y_t^2)-12\la_8(2g_{_{B-L}}^2+g_2^2)\\
\eeas
\item {\bf Coefficients with $\Phi^2 \Delta^2$}
\beas
{16\pi^2}\frac{d}{dt}\la_9 &=& \la_9\Big(20\la_1+8\la_3+16\la_5+8\la_6+6\la_7+4\la_9+6Y_t^2+4\Tr\big[(Y^M)^2\big]-6g_{_{B-L}}^2-21g_2^2\Big)\nonumber\\
&&+6g_2^4+16(\la_{10}^2+\la_{11}^2)+\la_{12}(8\la_1+\la_{12})+3\la_{13}^2+12\la_{14}^2\nonumber\\
&& +8\la_{12}\la_3+48\la_{10}\la_4+\la_{12}(6\la_5+8\la_6+3\la_7)\\
{16\pi^2}\frac{d}{dt}\la_{10} &=& \la_{10}\Big(4\la_1+4\la_{12}+48\la_2+16\la_3+16\la_4+16\la_5+8\la_6+6\la_7+8\la_9 \nonumber\\
&& +6Y_t^2+4\Tr\big[(Y^M)^2\big]-6g_{_{B-L}}^2-21g_2^2\Big)-3\la_{13}(\la_{14}+\la_{15})+12\la_4\la_9\\
{16\pi^2}\frac{d}{dt}\la_{11} &=& \la_{11}\Big(4\la_1+4\la_{12}-48\la_2+16\la_3+16\la_5+8\la_6-6\la_7\nonumber \\
&& +8\la_9+6Y_t^2+4\Tr\big[(Y^M)^2\big]-6g_{_{B-L}}^2-21g_2^2\Big)\\
{16\pi^2}\frac{d}{dt}\la_{12} &=& \la_{12}\Big(4\la_1+4\la_{12}-8\la_3+4\la_5-8\la_6+8\la_9+6Y_t^2\nonumber\\
&& 4\Tr\big[(Y^M)^2\big]-6g_{_{B-L}}^2-21g_2^2\Big)-12(\la_{14}^2-\la_{15}^2)\\
{16\pi^2}\frac{d}{dt}\la_{13} &=& \la_{13}\Big(4\la_1+4\la_{12}+8\la_3+2\la_7+8\la_8+8\la_9+3Y_t^2+\Tr\big[(Y^M)^2\big]\nonumber\\
&& -6g_{_{B-L}}^2-21g_2^2\Big)+\big(8\la_4+16\la_{10}\big)\big(\la_{14}+\la_{15}\big)\\
{16\pi^2}\frac{d}{dt}\la_{14} &=& \la_{14}\Big( 4\la_1-4\la_{12}+2\la_7+8\la_9+6Y_t^2+4\Tr\big[(Y^M)^2\big]-6g_{_{B-L}}^2-21g_2^2\Big)\nonumber\\
&& +4\la_{13}(\la_4+2\la_{10})+8\la_{15}(2\la_2+\la_8)\\
{16\pi^2}\frac{d}{dt}\la_{15} &=& \la_{15}\Big(4\la_1+12\la_{12}+2\la_7+8\la_9+6Y_t^2+4\Tr\big[(Y^M)^2\big]-6g_{_{B-L}}^2-21g_2^2\Big)\nonumber\\
&& +4\la_{13}(\la_4+4\la_{10})+8\la_{14}(2\la_2+\la_8)
\eeas

\end{itemize}

\subsection{LR Model with Doublet Scalars} \label{app:RG_LR_doublet}

\subsubsection*{Gauge RG Equations}
\beas
{16\pi^2}\frac{d}{dt}g_3 &=&  {g_3^3} \bigg(-7\bigg)\\
{16\pi^2}\frac{d}{dt}g_{2} &=&  {g_{2}^3} \bigg(-\frac{17}{6}\bigg)\\
{16\pi^2}\frac{d}{dt}g_{_{B-L}}  &=& {g^3_{B-L}} \big( 3 \big)
\eeas
Note that in our case $g_{_{2L}} = g_{_{2R}} = g_{_{2}}$. 

\subsubsection*{Fermion RG Equations}
\bes
{64 \pi^2}\frac{d}{dt}Y_{t} = \bigg(-\frac{2}{9} g_{_{B-L}}^2 - 9 g_2^2-32 g_3^2\bigg)Y_t + 7 Y_t^3
\ees


\subsubsection*{Scalar RG Equations}

\begin{itemize}
 \item {\bf Coefficients with $\Phi^4$}
 
 \beas
 {128\pi^2}\frac{d}{dt}\la_1 &=& \la_1\bigg(-72 g_2^2+256\big(\la_1+\la_2-\la_3\big)+24 Y_t^2\bigg)\nonumber\\
                            &+& 1024(\la_1^2+\la_2^2)+32\beta_1^2+8f_1^2+9g_2^4-12Y-Y_t^4 \\
{512\pi^2}\frac{d}{dt}\la_2 &=& \la_2\bigg(-288g_2^2+768\la_1+3072\la_2+1024\la_3+96Y_t^2\bigg)-8 f_1^2+3g_2^4-3Y_t^4 \nonumber \\
{256\pi^2}\frac{d}{dt}\la_3 &=&  \la_3\bigg(-144g_2^2-384\la_1-512\la_2-1536\la_3+48Y_t^2 \bigg)+ 4 f_1^2-3g_2^4-3Y_t^4 \nonumber 
\eeas

\item {\bf Coefficients with $H_{L/R}^4$}

\beas
{512\pi^2}\frac{d}{dt}\kappa_1 &=& \kappa_1\big(-96g_{_{B-L}}^2-144g_2^2+576\kappa_1+384\kappa_2\big)\nonumber \\
                               &+& 192\kappa_2^2+256\beta_1^2+128f_1^2+24g_{B-l}^4+12g_{_{B-L}}^2g_2^2+9g_2^4\\
{512\pi^2}\frac{d}{dt}\kappa_2 &=& \kappa_2\big(-96g_{_{B-L}}^2-144g_2^2+512\kappa_1+384\kappa_2\big)+128f_1^2+12g_{_{B-L}}^2g_2^2+9g_2^4\nonumber
\eeas

\item 

{\bf Coefficeients with $\Phi^2 H_{L/R}^2$}

\beas
{256\pi^2}\frac{d}{dt}\beta_1 &=&-4\beta_1\bigg[-8\beta_1+6g_{_{B-L}}^2+27g_2^2-2\big(20\kappa_1+4\kappa_2+40\la_1+32\la_2-32\la_3+
3Y_t^2\big) \bigg]\\&& + 24f_1^2+9g_2^4 \nonumber \\
{256\pi^2}\frac{d}{dt}f_1 &=& f_1 \bigg( 16\beta_1-6g_{_{B-L}}^2-27g_2^2+8(\kappa_1+\kappa_2)+16(\la_1-4\la_2)+64\la_3+6Y_t^2 \bigg)
\eeas
\end{itemize}


\bibliography{vacuum_stability}

\end{document}